# Sandro Caranzano[#] and Mariateresa Crosta[*]

[#]*Centro Studi Herakles, Torino*
[*]*Isitituto Nazionale di Astrofisica-Osservatorio Astrofisico di Torino*

# La data di fondazione di *Augusta Taurinorum ex sole*. Propaganda augustea e il ruolo dell'Astronomia[1]

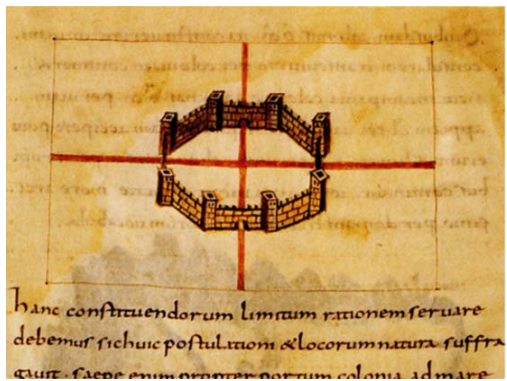

# INDICE

## Il Natale della città e il problema della cronologia di *Augusta Taurinorum*

La data di fondazione di *Augusta Taurinorum* non è nota; come in molti altri casi non disponiamo di testimonianze letterarie ed epigrafiche che permettano di inquadrare la deduzione della città nel più ampio contesto storico che coinvolse la Cisalpina nord-occidentale nel periodo di transizione tra la repubblica e il principato.

La quasi totalità degli studiosi ha accantonato l'ipotesi avanzata a partire dal XIX sec. di una doppia deduzione (prima come municipio sotto Giulio Cesare, poi come colonia sotto Augusto, dopo il 27 a.C.), una teoria fondata sul ritrovamento di iscrizioni con la dicitura *Iulia Augusta Taurinorum* alternate ad altre in cui ricorre il più semplice *Augusta Taurinorum*[2]*,* oltre che sulla sporadica citazione dei *quattuorviri* con potestà edilizia (una magistratura tipica dei *municipia*) in vece dei *duoviri* tipici delle colonie[3].

Per risolvere l'incongruenza si è osservato che il polimorfismo della titolazione urbica non sembra seguire una rigida scansione cronologica quanto piuttosto essere conseguenza della volontà di chi scriveva di indicare più o meno per esteso il nome della città[4], mentre l'utilizzo improprio del titolo quattuorvirale nelle colonie sembra spiegabile in qualità di variazione lessicale conseguente al fatto che i suoi membri appartenevano a un collegio magistratuale composto di quattro individui[5].

D'altra parte, dal punto di vista archeologico, i ripetuti scavi di emergenza nel centro storico e nelle sue adiacenze, motivati dai lavori pubblici, dalla costruzione di parcheggi e delle infrastrutture connesse allo svolgimento delle Olimpiadi invernali 2006, non hanno offerto evidenza dell'esistenza in loco di un *castrum stativum* cesariano.

Evanescenti sembrano anche essere i resti monumentali di età repubblicana.

I sondaggi eseguiti in piazza Castello presso Palazzo Madama hanno permesso di raccogliere alcuni esemplari di ceramica in corrispondenza delle fosse di fondazione della cinta romana che inquadrano il cantiere di costruzione delle mura orientali attorno al 70 d.C. (nei primi anni del regno di Vespasiano), vale a dire in un periodo più avanzato di quando si immaginasse[6].

---

[2] *CIL* V, 6954; *CIL* V, 7047. Cfr. PROMIS, 1869, pp. 58, 63 e 69; CRESCI MARRONE, 1977, pp. 137-139. Secondo le prime teorie, nata inizialmente con il nome di *Iulia* ad opera di Cesare (tra il 49 e il 44 a.C.) oppure di Ottaviano (dopo il 42-41 a.C.), la città sarebbe stata rifondata dopo il 27 a.C. da Ottaviano Augusto ricevendo l'attributo di 'augusta'(cfr. PROMIS, 1869, INAUDI, 1976, pp. 381-398). Sappiamo però che anche le colonie dedotte dopo il 27 a.C. potevano essere appellate *Iuliae Augustae*. Un'interessante sintesi sui problemi relativi alla cronologia di fondazione della città si trova in MASCI, 2012, pp. 63-78.

[3] *CIL* V, 7028; *CIL* V 7034*; CIL* V, 7037, *CIL* V, 7039. Inizialmente interpretate come testimonianza di una doppia fondazione della città, le epigrafi sono state recentemente spiegate come la prova di un'istituzione municipale promossa dagli indigeni Taurini al ricevimento della cittadinanza romana nel 49 a.C., prima però che venisse fondata la colonia vera e propria (Cfr. CRESCI MARRONE, 1977, pp. 139-141).

[4] PACI, 1998, p. 111.

[5] *Ivi*., pp. 121-122. Inoltre cfr. CRESCI MARRONE, 1977, cit., p. 149.

[6] BRECCIAROLI TABORELLI, GABUCCI, 2006, pp. 252-256. Già nel 1981, in occasione degli scavi per la posa di tubature allo sbocco di via Garibaldi su piazza Castello fu segnalata l'insolita presenza di due frammenti di ceramica sigillata di età neroniana al di sotto del basolato in pietra del decumano massimo. Nell'occasione si

Anche in occasione di un sondaggio effettuato nel cortile di uno stabile di via Cesare Battisti è stato possibile individuare i resti di una *domus* del secondo quarto del I sec d.C. che, dopo una ventina di anni di vita, venne affiancata dai muraglioni difensivi del lato meridionale della città, a conferma del fatto che anche questi ultimi furono edificati nel terzo quarto del I sec. d.C.[7].

Leggermente differente è la situazione delle mura urbiche settentrionali; qui, una stratigrafia rilevata nel punto nel quale le mura intercettano la *porticus* del teatro romano dimostrano che esse furono costruite con un po' di anticipo rispetto a quelle orientali[8], dovendosi inquadrare nei primi decenni dopo la nascita di Cristo.

Il ritrovamento più spettacolare deve tuttavia considerarsi quello effettuato in piazza Castello, immediatamente a ovest del muro di cinta romano, dove quattro anfore disposte ai vertici di un quadrangolo (5,30 x 4,15 x 5,60 x 4,50 m) delimitavano un rogo rituale di consacrazione. La foggia delle anfore di tipologia *Péliche* 46 rimanda al terzo quarto del I sec a.C. (di fatto, all'età di Claudio e di Nerone); ne consegue che il rito inaugurale della cinta urbica avvenne nel corso della seconda metà del I sec. d.C.[9].

È qui bene sottolineare che il rito di consacrazione scoperto in prossimità delle mura dovrà riferirsi esclusivamente all'inaugurazione delle medesime, atto che non coincide necessariamente con la fondazione della colonia che nel nostro caso, senza dubbio, avvenne in precedenza.

Come si vede, dal punto di vista infrastrutturale, la colonia taurinense sembra aver assistito ad un'attività costruttiva degli edifici pubblici lenta e progressiva, in armonia d'altronde con quanto registrato in altre fondazioni provinciali, anche in considerazione dell'impegno tecnico e finanziario necessario[10] (figg. 3-4).

Le interessantissime scoperte archeologiche fanno, insomma, luce sui tempi e sui modi della monumentalizzazione urbica, ma non sembrano ancora in grado di offrire una soluzione definitiva al problema della cronologia di fondazione della colonia.

Dal punto di vista storiografico è degno di nota che quando nel 58 a.C. Gaio Giulio Cesare attraversò a marce forzate la valle di Susa per intervenire contro gli *Helvetii*, all'inizio delle Guerre galliche, il condottiero subì azioni di disturbo da parte delle tribù insediate nelle valli alpine, ma pur avendo percorso la pianura torinese in direzione *Ocelum* (un centro indigeno con ogni probabilità ubicato nei pressi delle Chiuse), nei Commentari non fa parola di Torino o dei Taurini[11].

---

segnalò l'assenza di reperti di altre epoche, ma venne sottolineato che i cocci avrebbero anche potuto depositarsi in occasione di una risistemazione o di un restauro condotti nell'antichità (FILIPPI, 1982, pp. 65-118).

[7] BRECCIAROLI TABORELLI, GABUCCI, 2006, pp. 243-245.
[8] *Ivi.*, pp. 247-251.
[9] *Ibid.*, pp. 244-246; BRECCIAROLI TABORELLI, PEJRANI BARICCO, Tracce di uno spazio sacrificale presso le mura di *Augusta Taurinorum*, in A. Carandini e R. Cappelli (a cura di) *Roma, Romolo, Remo e la fondazione della città*, cat. mostr., Roma 2000, pp. 281-282; BRECCIAROLI TABORELLI, PEJRANI BARICCO, OCCELLI, (2001), pp. 98-99; PEJRANI BARICCO, MAFFEIS, 2006, p. 20.
[10] FURGER, ISLER KERÉNYI *et al.*, 2001; LAUR-BELART, 1991; BOGLI, 1989; MOLLO MEZZENA, 1982, pp. 63-134; TORELLI, 1988, pp. 29-48.
[11] CAES. *de bello.* 1, 10.

Anche l'epigrafia è rara di testimonianze certamente attribuibili all'età repubblicana e i pochi documenti caratterizzati da una scrittura e da un formulario arcaico (tra l'altro non direttamente provenienti dal sito di Torino, ma dal suo territorio) sono stati spiegati come semplice effetto di fenomeni di conservatorismo grafico e stilistico[12].

Il recente ritrovamento presso Alpignano di un termine campestre databile al 21 a.C. nel quale viene esplicitamente citato il consolato di *M(arco) Lollio* ha riaperto il dibattito: pubblicato da G. Mennella[13] come una prova dell'avvenuta deduzione della colonia di *Augusta Taurinorum* negli anni immediatamente successivi al 27 a.C., il documento è stato per contro considerato da G. Masci una prova della mancanza, a quel tempo, di un riferimento amministrativo municipale o coloniale organizzato nelle vicinanze[14]; in sintesi, rovesciando il punto di vista, una prova che a quel tempo *Augusta Taurinorum* non era ancora stata fondata.

Un punto fisso sul quale tutti sono ormai d'accordo è che la titolatura di '*augusta*' attribuito alla città offre un termine *post quem* al 16 gennaio del 27 a.C., data nella quale il Senato di Roma attribuì tale titolo onorifico a Ottaviano.

Se poi passiamo ad analizzare i celebri itinerari incisi sui Vasi di Vicarello possiamo osservare che il toponimo di *Augusta Taurinorum* appare unicamente nel quarto della serie, dove è proposto un *itinerarium* di cui non conosciamo l'origine ma che fu stilato dopo il 13 a.C. in ragione del fatto che vi viene menzionata la prefettura delle Alpi Cozie (che prima di quella data non esisteva)[15]. Nei primi tre esemplari, che invece riportano un itinerario derivato da un modello scolpito a *Gades* tra il 24 e il 19 a.C., il nome della città è assente e la tappa viene indicata semplicemente facendo uso della dicitura *Taurinis*.

Se passiamo all'analisi dei testi degli scrittori antichi, nuovamente i riferimenti alla fase più antica della storia cittadina appaiono piuttosto evanescenti ed i primi autori che nominano esplicitamente *Augusta Taurinorum* (Plinio il Vecchio[16] e Svetonio[17]) si

---

[12] PACI, 1988, pp. 107-131.
[13] MENNELLA, 2012, pp. 387-394; MASCI, 2012, pp. 63-78; BARELLO, 2016, pp. 290-291.
[14] «La presenza di un *fundus* assegnato ad un privato potrebbe non implicare necessariamente l'esistenza di una colonia; la datazione consolare, non frequente in questa tipologia di iscrizioni, può anzi suggerire la mancanza di un contesto di riferimento amministrativo stabilito e organizzato, portando a credere che il proprietario del terreno abbia avvertito il bisogno di trovare un rimando altro cui ancorare la dichiarazione di proprietà». (MASCI, 2012, pp. 63-78). A tal proposito è bene ricordare che già nel 1997 G. Cresci Marrone aveva ipotizzato l'esistenza di un'iniziativa di urbanizzazione della popolazione taurina nel periodo immediatamente successivo alla concessione della cittadinanza romana della Transpadana del 47 a.C. «sollecitata a rafforzare con un moto sinecistico 'di aggregazione' le proprie precarie forme di insediamento forse già attive alla confluenza della Dora nel Po, dunque nel sito della Torino romana», un fatto questo che avrebbe dato vita ad una breve fase municipale senza che Roma dovesse necessariamente dedurre sul posto una colonia (cfr. CRESCI MARRONE, 1997, pp. 140-141).
[15] FRANCE, 2001; MASCI, 2012, pp. 66-67.
[16] PLIN. *nat. hist.* 3, 123. Plinio il Vecchio cita Torino nella prefazione del Libro 3 della *Naturalis Historia* e precisa di aver consultato una lista alfabetica delle città d'Italia redatta da Augusto. Plinio ricorda anche che a quel tempo il Po era navigabile fino a Torino mentre Polibio, due secoli prima, si limitava a ricordarne la navigabilità solo sino a *Vardacate "Industria"* (POLYB. 2, 16).
[17] Così in PLIN. *nat. hist.* 3, 123: *Transpadana appellatur ab eo regio undecima, tota in mediterraneo, cui marina*

situano in momento cronologico piuttosto avanzato e di scarsa utilità nel contesto del dibattito sulla cronologia di deduzione[18].

Nell'ambito della discussione sulla cronologia, merita una riflessione il fatto che nell'excursus riservato all'Italia antica dal geografo Strabone Torino non sia citata, laddove invece è menzionata *Augusta Praetoria*. Benché la data di composizione della *Geografia* sia oggetto di dibattito, la critica moderna è concorde nel ritenere che la sua redazione sia stata conclusa tra il 18 e il 24 d.C., anche in considerazione del fatto che al suo interno si fa riferimento ad alcuni eventi relativi al principato di Tiberio[19]. Il problema da porsi è tuttavia quello relativo alle fonti utilizzate: senza entrare nel complesso dibattito inerente alla presenza di due fasi di scrittura e alla loro cronologia (in ogni caso, la prima redazione della Geografia è stimata dai massimalisti non più addietro del 7 a.C.[20]), sembra chiaro che Strabone revisionò o completò l'opera alla veneranda età di circa ottant'anni, sfruttando diverse fonti scritte integrate dalle proprie personali esperienze di viaggio, come d'altronde dimostrato da alcuni anacronismi rilevabili nell'opera.

Il fatto che Strabone citi *Augusta Praetoria* e non *Augusta Taurinorum*[21] sembra difficile da giustificare con una svista dell'autore, tanto più che la colonia non sorgeva in una posizione remota ma all'interno dei confini dell'Italia antica. Si potrebbe proporre che la lacuna derivi dal fatto che il geografo utilizzò una fonte 'datata' e quindi non al corrente della sua fondazione, e che l'autore non sentì il bisogno o non ebbe modo di integrarla[22].

Vi è tuttavia un documento particolarmente significativo che lo scrittore avrebbe dovuto conoscere, vale a dire la cosiddetta Mappa di Agrippa, una monumentale carta del mondo conosciuto accompagnata da didascalie che fu esposta in Campo Marzio dopo

---

*cuncta fructuoso alveo inportat. oppida Vibi Forum, Segusio, coloniae ab Alpium radicibus Augusta Taurinorum — inde navigabili Pado — antiqua Ligurum stirpe, dein Salassorum Augusta Praetoria iuxta geminas Alpium fores, Graias atque Poeninas — his Poenos, Grais Herculem transisse memorant —, oppidum Eporedia Sibyllinis a populo Romano conditum iussis. Eporedias Galli bonos equorum domitores vocant*; TAC. *hist.* 2, 66: *Augustae Taurinorum, dum opificem quendam Batavus ut fraudatorem insectatur, legionarius ut hospitem tuetur, sui cuique commilitones adgregati a conviciis ad caedem transiere. et proelium atrox arsisset, ni duae praetoriae cohortes causam quartadecimanorum secutae his fiduciam et metum Batavis fecissent.*

[18] CRESCI MARRONE, 1997, p. 138 e 147.
[19] DUECK, 1999, pp. 467-468. Un'unica notazione di cronologia assoluta sembra presente quando Strabone, narrando delle operazioni militari di Tiberio e Druso contro i Carni e i Norici, dice distintamente che esse avvennero 33 anni prima, segno che la compilazione del paragrafo in oggetto era in atto nel 19 d.C. (cfr. FABRICIUS, 1717, pp. 3-4).
[20] PAIS, 1908, p. 401.
[21] STR. *Geo.* 4, 6 e 5, 1. *Augusta Taurinorum* appare naturalmente nella *Tabula Peutingeriana* compilata nella tarda antichità partendo – secondo alcuni – dalla Tavola di Agrippa, ma trattandosi di una mappa destinata ai viaggiatori è naturale che, a quel tempo, eventuali lacune fossero state opportunamente integrate (cfr. L. CRACCO RUGGINI, *Torino tra antichità* cit., pp. 11-12).
[22] La problematica viene affrontata in uno studio sulla cronologia di *Segusio* dove la mancata citazione del centro valsusino da parte di Strabone viene spiegato con l'utilizzo di una fonte stilata poco dopo la stipula del *foedus* con Roma, dunque negli anni immediatamente successivi il 13 a.C. (cfr. VOTA, 2003, pp. 25-26).

la morte del genero di Augusto nel 12 a.C.[23], probabilmente a partire dal 7 a.C. o poco dopo.

Non vi è dubbio che, in qualità di geografo, Strabone abbia consultato la mappa[24], ma è curioso che nonostante ciò egli non abbia apportato alcuna correzione o integrazione alla sua descrizione dell'Italia antica. Ciò potrebbe essere conseguenza del fatto che l'*orbis pictus* esposto da Augusto nella *Porticus Vipsania* non ne riportava il nome, un fatto tuttavia veramente anomalo, tanto più che si trattava di una fondazione ubicata in una delle regioni augustee dell'Italia romana. Tale lacuna si spiegherebbe, però, ammettendo che, a quel tempo, *Augusta Taurinorum* non esistesse ancora, oppure che, trattandosi di una fondazione recentissima, non fosse citata nella corografia che gli artefici della Mappa di Agrippa utilizzarono come modello.

Naturalmente siamo di fronte ad un argomento *ex silenzio* ma se le cose andarono così ne deriverebbe una datazione di *Augusta Taurinorum* non precedente alla metà del secondo decennio a.C.

L'utilità di ricostruire con maggior precisione la data di fondazione di *Augusta Taurinorum* riguarda quel ristretto numero di anni che va dal 27 a.C. all'età di Cristo, vale a dire quel periodo storico che va dalla concessione del titolo di *augusto* ad Ottaviano al momento nel quale la città viene esplicitamente citata dagli scrittori antichi.

Per risolvere il problema si è tentato un approccio basato sulle corrispondenze esistenti tra il calendario solare e l'orientamento inaugurale delle città etrusco-romane che sappiamo essere stato rispettato per lungo tempo a Roma e nelle sue province; si tratta pertanto di un'indagine interdisciplinare che necessita dell'ausilio dell'astronomia.

**Premessa metodologica. L'orientamento *ex sole***

Il recente sviluppo di una disciplina acheoastronomica ha posto le basi per uno studio più accurato del rapporto esistente tra architettura, urbanistica e astronomia antica[25]. Che la fondazione di una città 'inaugurata' fosse sottoposta a precise prescrizioni sacre è ribadito da Sesto Pompeo Festo: *rituales nominantur Etruscorum libri, in quibus praescriptum est quo rito condantur urbes, arae aedes sacrentur, qua sanctitate muri, quo iure portae, quomodo tribus, curiae centuriae distribuantur, exercitus constituantur,*

---

[23] DILKE, 1987, pp. 207-209. La descrizione del mondo avrebbe potuto comporsi di più tavole quadrangolari affiancate e non avere – come spesso si pensa – forma circolare ad imitazione del globo. Essa fu esposta presso la *Porticus Vipsania* ma il suo aspetto preciso non è noto (cfr. TIERNEY, 1962, pp. 151-166; DUECK, 1999, pp. 467-468).

[24] Strabone fu a Roma il 7 a.C. e nella Geografia cita spesso una «corografia» o «mappa corografica»; benché non faccia mai esplicito riferimento ad Agrippa – forse per non sminuire il peso del principe che aveva promosso il completamento del portico – la prevalenza degli studiosi è convinta che l'espressione si riferisca proprio alla mappa dell'*oikoumene* esposta in Campo Marzio (cfr. TIERNEY, 1962, pp. 467-468).

[25] MAGLI, 2007, pp. 63-71.

*ordinentur, ceteraque eiusmodi ad bellum ac pacem pertinentia*[26].

Igino Gromatico sottolinea con chiarezza il collegamento sussistente tra il decumano e il corso del sole: *Constituti enim limites non sine mundi ratione, quoniam decumani secundum solis decursum diriguntur, kardines a poli axe. Unde primum haec ratio mensurae constituta ab Etruscorum aruspicum disciplina*[27]. E più oltre: *quare non omnis agrorum mensura in orientem potius quam in occidentem spectat. In orientem sicut aedes sacrae. Nam antiqui architecti in occidentem templa recte spectare scripserunt: postea placuit omnem religionem eo convertere, ex qua parte caeli terra inluminatur. sic et limites in orientem constituuntur*[28].

Di particolare interesse è, infine, un passo di Frontino in cui viene rimarcata la relazione esistente tra l'orientamento della città e la volta celeste, tra la gromatica e la *disciplina* etrusca[29]: *limitum prima origo, sicut Varro descripsit, a[d] disciplina[m] Etrusca[m]; quod aruspices orbem terrarum in duas partes diviserunt, dextra appellaverunt [quae] septentrioni subiaceret, sinistram quae ad meridianum terrae esset, ab orientem ad occasum, quod eo sol et luna spectaret, sicut quidam carpiunt architecti delubra in occidentem recte spectare scripserunt. Aruspices altera[m] linea[m] a septentrionem ad meridianum diviserunt terram, et a me[ri]dia[no] ultra antica, citra postica nominaverunt.*

*Ab hoc fundamento maiores nostri in agrorum mensura videntur constituisse rationem. Primum duo limites duxerunt; unum ab oriente in occasum, quem vocaverunt decimanum; alterum a meridiano ad septentrionem, quem cardinem appellaverunt. Decimanus autem dividebat agrum dextera et sinistra, cardo citra e ultra.*

Benché il passo di Frontino non scenda nei dettagli e faccia un generico riferimento alle relazioni sussistenti tra la *limitatio* e l'etrusca *disciplina*, è evidente come implicitamente chiarisce che il decumano – e non il cardo – veniva a costituire l'asse generatore dell'impianto urbano. Inoltre, Frontino indica chiaramente che esso veniva tracciato proprio da oriente verso occidente (e non in senso opposto).

---

[26] Fest. 285.

[27] Hyg. Grom. *const. lim.*: «I limiti non sono stabiliti senza che si tenga conto della misura dell'universo, perché i decumani sono orientati secondo il corso del sole e i cardini secondo quello dei poli. E tali principi di *mensura* sono stati stabiliti all'inizio dalla *disciplina* aruspicale etrusca».

[28] *Ivi,* 4: «Per cui la misurazione dei campi è orientata ad oriente piuttosto che ad occidente. Infatti nei tempi antichi gli architetti scrissero a ragione che i templi devono essere rivolti a ponente; in seguito si preferì volgere tutto ciò che ha a che fare con la sfera religiosa da quella parte del cielo da cui la terra viene illuminata. Di conseguenza anche i confini vengono fissati in direzione dell'oriente».

[29] Front. *de lim.*: «L'origine prima dei *limites* – come d'altronde è stato esposto da Varrone – deriva dalla *disciplina* etrusca: 'che gli aruspici divisero l'orbe della Terra in due parti, denominando destra quella che si trova a settentrione, sinistra quella che si trova a meridione, dall'Oriente all'Occaso, secondo il corso del sole e della luna (sicché alcuni architetti hanno scritto, a ragione, che le are sacre devono essere poste ad occidente). Gli aruspici poi, con un'altra linea, hanno diviso lo spazio da nord a sud e hanno dominato *antica* la parte che si pone al di là di tale linea, *postica* quella che sta al di qua. Su tali basi, i nostri padri hanno fondato i principi dell'agrimensura. Per iniziare essi tracciavano due limiti: uno da oriente ad occidente denominato *decumanus*; un altro da sud verso nord denominato *cardo*. Il decumano divide il terreno tra parte *dextera* e parte *sinistra*, il *cardo* lo divide tra parte *citra* e parte *ultra*».

Un passo particolarmente importante, che sino ad oggi non sembra essere stato compreso in tutte le sfumature[30], è infine quello nel quale Igino allude esplicitamente al metodo con il quale si era usi tracciare i *limites* di una colonia o di un *castrum* al momento della fondazione. Egli infatti scrive: *multi ignorantes mundi rationem solem sunt secuti, hoc est ortum occasum, quod is semel ferramento comprehendi non potest. Quid ergo? Posita auspicaliter groma, ipso forte conditore praesente, proximum vero ortum comprehenderunt, et in utramque partem limites miserunt, quibus cardo in horam sextam non convenerit*[31], vale a dire: «Molti, ignorando la conformazione del cosmo, si sono adeguati al corso del sole, ovvero al suo levare e al suo calare, cosa che non può essere delimitata una volta per tutte con uno strumento. Quale conseguenza tutto ciò comporta? Dopo aver preso gli auspici e aver collocato la groma, eventualmente alla presenza dello stesso fondatore della città, essi [sc. i gromatici] hanno definito con precisione il punto presso il quale sorge il Sole, hanno quindi definito i *limites* da entrambe le parti [la *citeriore* e la *posteriore*] con le quali il cardo non verrà a coincidere che all'ora sesta»[32].

Igino intende insomma chiarire che i gromatici stabilivano l'orientamento del decumano osservando il corso del sole la cui posizione, cambiando di giorno in giorno, non era determinabile a priori ma doveva essere misurata per mezzo di uno strumento, cioé la groma. Al sorgere del sole venivano dunque tracciati i *limites* della città, vale a dire il decumano (allineato con il punto di levata del sole) e, subito dopo, il cardo che divideva la città in due parti distinte, delle quali la prima posta ad occidente e la seconda ad oriente (vale a dire la *pars citra* e la *pars ultra*: cfr. nota 27).

Con tale affermazione, Igino esplicita chiaramente che il tracciamento del cardo avveniva contestualmente a quello del decumano. Per far ciò non era necessario far uso di uno strumento astronomico come lo gnomone, poiché era sufficiente impiegare i bracci della groma che, come sappiamo, erano disposti a croce e formavano un angolo di novanta gradi.

Con un giro di parole apparentemente oscuro per la sua sinteticità, Igino lascia intendere che solamente a mezzogiorno l'ombra proiettata da uno gnomone sarebbe andata ad allinearsi con il cardo così tracciato (come d'altronde è naturale sia), dimostrando la correttezza del lavoro svolto secondo il principio enunciato. È evidente che la trattazione fu rivolta ad un pubblico di tecnici e necessita di una certa pazienza per essere compresa, proprio perché dà per scontate diverse cose, mentre il linguaggio dei gromatici suona spesso oscuro.

Interpretato correttamente, il passo di Igino ci offre una preziosissima ed unica descrizione delle modalità con le quali si svolgeva la fondazione di una città alla levata del sole e chiarisce, al contempo, che il gromatico era affiancato dal *conditor* che normalmente era nominato e inviato sul posto dal Senato di Roma.

---

[30] DILKE, 1987, p. 25.
[31] HYG. GROM. *const. lim.* 4.
[32] Nell'antichità per 'ora sesta' si intendeva sia il mezzogiorno astronomico, sia lo spazio di tempo a seguire (sino a tre ore) con una certa indeterminatezza. Igino asserisce pertanto la verità, ma si può supporre la fonte a cui attinse (e che egli comprese solo superficialmente) circoscrivesse l'esempio limitatamente al mezzogiorno delle date equinoziali e che l'autore, proprio perché *ignorans mundi rationem,* non abbia in realtà compreso in profondità il trattato astronomico che consultò.

Recentemente, la verifica dell'orientamento solare equinoziale e solstiziale di diversi *castra* e città romani ha rafforzato l'idea che la direzione del decumano fosse stabilita osservando il punto di levata del sole all'orizzonte locale il giorno della fondazione, tracciando il decumano lungo la linea che univa tale punto con l'*umbilicus* nel quale veniva effettuata la misurazione[33]. È questo il caso di *Lugdunum* (Lione), città fondata nel 43 a.C. ed allineata con il sole nascente il 1 agosto (cioè con la ricorrenza celtica del Lughnasadh)[34], della *Colonia Ara Augusta Agrippinensis* (Colonia, in Germania)[35], orientata con la levata del sole il 23 settembre, giorno natale di Augusto e, come si vedrà, di *Augusta Praetoria* (Aosta).

Si veniva così a determinare una consacrazione della città che si manifestava fisicamente e visivamente sulla rete stradale e sugli edifici sacri: in particolare, in occasione del genetliaco urbico, il sole, trovandosi in posizione perfettamente ortogonale rispetto ai *cardines*, sarebbe sorto illuminando le strade e i templi con un indubbio effetto scenografico, percorrendo poi nel corso della mattinata un arco nel cielo sino alla sua Culminazione. Il rapporto l'orientamento degli edifici sacri e il sorgere del sole è accennato anche in Vitruvio, che ricorda il rapporto emozionale che si veniva a creare tra i fedeli quando il sole, sorgendo alle spalle dei templi opportunamente orientati, si presentava davanti agli altari simile all'effigie della divinità[36].

Si tratta di un procedimento non molto differente da quello impiegato nel periodo medievale per celebrare la ricorrenza di determinate feste religiose in svariate chiese e santuari, ove finestre, rosoni e muri erano deviati di qualche grado rispetto al resto dell'edificio per ottenere particolari effetti scenici; il caso piuttosto noto della basilica romana di Sant'Aquilino a Milano (nella quale al solstizio di inverno il sole illumina la nicchia mosaicata ad est dell'ottogono illuminando la figura di *Christos-Helios*) conferma che l'utilizzo di alcuni *edificia* in qualità di *horologia* è di fatto un'eredità del

---

[33] A sostegno di tale proposta si è calcolato che su 38 città romane dell'Italia antica ben 35 presentano un orientamento nell'arco di orizzonte percorso dal sole alla sua levata nel corso dell'anno; inoltre, due di esse, Verona e Vicenza, sono deliberatamente orientate sulla linea del solstizio estivo e altre nove in prossimità della linea del solstizio invernale (cfr. MAGLI, 2007, pp. 63-71). Di parere differente J. Le Gall che interpreta il passo di Igino con una differente traduzione e ne consegue la prova che l'inaugurazione non avveniva all'alba (cfr. J. LE GALL, 1975, pp. 287-320).

Sull'argomento si vedano i casi più significativi: AVENI, ROMANO, 1994, pp. 545-563; LANCIANO, VIRGILI, 2016, pp. 249-255. Inoltre: AVENI, CAPONE, 1985, pp. 5-16; BROWN, 1980; DECRAMER 2011; LE GALL, 1975, pp. 287-320; TORELLI, 1966; HAWKINS, 1985.

[34] ESPINOSA et al., 2014, pp. 107-119.

[35] ID., 2016, pp. 234-239.

[36] VITR. *De. Arch.* IV, 5: «L'orientazione dei templi sarà da stabilire così che, se nessuna ragione lo impedisca e se vi sia libertà di scelta, il tempio e il simulacro del dio collocato nella cella guardino ad occidente. Così quelli che si presentano all'ara per il sacrificio si volteranno verso oriente e verso il simulacro nella cella, e offrendo i loro voti guarderanno il tempio e il cielo orientale: sembrerà che l'effigie del dio, quasi in persona sorgente, guardi dall'oriente i fedeli che supplicano e sacrificano».

mondo antico[37].

Naturalmente, tale approccio pratico non si opponeva affatto alla necessità di armonizzare l'orientamento della *limitatio* con la natura orografica e idrografica del territorio da assegnare[38]; probabilmente, il giorno di inaugurazione veniva scelto sulla base di considerazioni politiche e cultuali, partendo da una rosa di festività o di ricorrenze sufficientemente ampia. Tra di esse veniva forse scelta quella che meglio si armonizzava con le esigenze di sfruttamento agricolo ed economico del territorio circostante e dopo aver ben considerato le vie di comunicazione preesistenti e quelle in via di realizzazione. D'altronde, è bene ricordare che non sempre l'orientamento della centuriazione e quello della città coincidevano; al di là delle stratificazioni storiche, in molti casi non è da escludere che le divergenze tra orientamento dell'*urbs* e quella dei campi centuriati sia da spiegarsi nell'ottica di una disparità tra le esigenze sacrali e quelle connesse alla fruizione ottimale del territorio (drenaggio delle acque, esposizione ai venti e al sole, vie di comunicazione etc.)[39].

Si è scelto, pertanto, di studiare l'orientamento astronomico di Torino basandosi su algoritmi astronomici per un Sole Vero, tenendo conto anche dei limiti di un'osservazione ad occhio nudo e della conformazione geografica locale.

Considerando quale periodo significativo quello compreso tra il 27 a.C. (data nella quale Ottaviano fu elevato dal Senato al rango di Augusto) e il 14 d.C. (data della sua morte, cfr. tabella β in appendice), determinato con precisione l'orientamento del decumano (attuale via Garibaldi), impostate la latitudine e la longitudine locale, si è provveduto a ricostruire il cielo dell'antichità nel momento della levata del Sole sul profilo della collina torinese (pertanto, tenendo anche conto della sua elevazione). Definito come asse di osservazione quello che unisce l'*umbilicus* (presso il quale stazionava l'àugure) con il punto di levata del sole, si è verificata l'esistenza di un 'tema astrologico augurale' della città. In sostanza, dato un moto locale del Sole, si sono determinate le diverse date dell'antichità nelle quali si verificò una coincidenza tra l'azimut del Sole alla sua levata a Torino e l'orientamento del decumano.

Individuati i potenziali giorni di fondazione nell'intervallo di anni sopraindicato, si è quindi consultato il calendario romano per verificare se le date coincidessero con una specifica ricorrenza o festività, ponendo attenzione a quei giorni che assumevano un particolare significato.

---

[37] SPINAZZÈ, 2016, pp. 455-463; DALLAS, 2015, pp. 213-222. Nello specifico di Milano cfr. NERI, 2015, p. 33.
[38] LE GALL, 1975, p. 306.
[39] Non è detto che una regola così rigorosa fosse rispettata anche nella fondazione dei *castra* militari per i quali la disciplina gromatica consiglia una soluzione semplificata, con l'asse del *cardo* orientato in senso nord-sud (cfr. DILKE, 1987, pp. 41-42). Infine non siamo certi del fatto che lo stesso scrupolo nel rispettare la tradizione etrusco-romana sia stato costante nel corso del tempo. Ottaviano Augusto si distinse per l'interesse e il rispetto nei confronti delle tradizioni e dei riti aviti: *Nonnulla etiam ex antiquis caerimonis paulatim abolit restituit* (SVET. *Aug.* 31).

## Augusto 'archeologo' e il mito romuleo

L'applicazione degli antichi rituali della *disciplina* etrusca per la fondazione di una città è attestata nelle deduzioni di età repubblicana e non vi è dubbio che continuò ad essere applicata anche in seguito. A titolo di premessa, è forse opportuno ricordare come sotto il principato augusteo si manifestò, in modo esplicito, un rinnovato interesse per la tradizione della Roma delle origini e, in particolare, verso l'atto del *condere* 'fondare', ritenuto di particolare importanza simbolica e propagandistica. La letteratura e l'arte dell'età augustea, d'altronde, non risparmiano riferimenti espliciti al mito di fondazione attuato da Romolo e Remo, e Augusto non fece mai segreto di considerare se stesso il «novello Romolo»[40]; inoltre, egli curò con particolare attenzione il rispetto della *disciplina* greco-romana nonché dei *rituales*. Svetonio, d'altronde, chiarisce: *Nonnulla etiam ex antiquis caerimoniis paulatim abolita restituit*[41].

Che Augusto fosse interessato a Romolo e alle origini di Roma non è un segreto, tanto più che scelse come primitiva residenza (nucleo di partenza per una futura *domus* più ampia ed articolata) la casa del retore Ortensio che sorgeva quasi al vertice delle *Scalae Caci*, la cui porta fronteggiava l'antica capanna di Romolo e l'*auguraculum* presso il quale erano stati presi gli auspici per la fondazione della Capitale[42].

Non lontano dalla *domus* di Augusto sul Palatino, gli scavi condotti nel 1886 da P. Rosa permisero di recuperare alcuni piloncini arcaicizzanti di peperino sui quali sono riportati i nomi di *Marspiter, Remureine, Anabestas* e *Ferter Resius*. Le colonnine sembrano aver costituito dei segnacoli sormontati da immagini ed erano originariamente disposte ai vertici di un'area sacra quadrangolare edificata nel punto in cui si credeva di identificare lo scenario del fatale scontro tra Romolo e Remo ai tempi mitici della fondazione della città[43]. Il complesso sacro sorgeva nei pressi del Tempio di Giove Statore e si data all'età di Augusto. Fu l'imperatore, in effetti, a reintrodurre lo *ius fetiale* che regolava le dichiarazioni di guerra ed i trattati di pace e la cui composizione veniva fatta risalire al re degli Equi *Ferter Resius*[44].

L'interesse di Augusto per la fondazione di nuove città e colonie, nonché il rispetto per gli antichi rituali etrusco-italici, traspaiono con chiarezza da molteplici documenti.

Gaio Tranquillo Svetonio riferisce che il principe «ripopolò l'Italia con 28 colonie che egli stesso aveva provveduto a fondare, dotandole di moltissime opere pubbliche e di rendite garantendo loro, almeno in parte, pari dignità per diritto e onore alla stessa Urbe»[45]. Tra di esse, la maggioranza degli studiosi, annovera anche *Augusta*

---

[40] BIANCHI, 2008, pp. 7-55.
[41] SVET. *Aug.* 31.
[42] CARANDINI, 2008, pp. 30-33.
[43] Sembrano provarlo gli evidenti riferimenti a Remo/*Remoria* e la presenza del verbo greco ἀναβαίνω («salire», «arrampicarsi»). Cfr. TOMEI, 1997, pp. 36-37.
[44] TOMEI, 1993, pp. 621-659.
[45] SVET. *Aug.* 46. È oggetto di discussione la quasi totale assenza di nominativi appartenenti a veterani di guerra, ex soldati e legionari nel corpus epigrafico di *Augusta Taurinorum* (cfr. CRESCI MARRONE, 1997, pp. 148-149).

*Taurinorum*[46].

Anche nelle *Res Gestae,* compilate dall'imperatore in persona e poi affidate alle Vestali assieme al suo testamento, Augusto ricorda di aver condotto circa 500.000 cittadini romani sotto le insegne delle legioni e specifica che oltre 300.000 di loro, al congedo, ottennero come premio un'assegnazione di terra o del denaro, venendo ricondotti presso il municipio di provenienza o insediati in una delle colonie appena dedotte[47].

Al ventottesimo paragrafo, Augusto ritorna sull'argomento asserendo di avere fondato 28 colonie *celeberrimae et frequentatissimae* entro i confini dell'Italia antica*: Italia autem XXVIII [colo]nias, quae vivo me celeberrimae et frequentissimae fuerunt, me [auctore] deductas habet*[48].

Vi è poi un'ulteriore testimonianza degna di approfondimento che dimostra l'importanza sacrale e simbolica attribuita da Augusto al rito del *condere*: nel dizionario enciclopedico di Sesto Pompeo Festo, alla voce *Roma Quadrata,* lo scrittore ricorda un altare fatto costruire da Augusto di fronte al Tempio di Apollo sul Palatino (vale a dire nelle immediate vicinanze della propria *domus*) e specifica che «qui è posto ciò che di buon auspicio si suole usare nella fondazione di un città»[49].

L'altare della *Roma Quadrata* sorgeva al centro di un porticato che ospitava la cosiddetta *Silva Apollinis*, vale a dire un boschetto di allori sacri sospeso sulla sottostante valle Murcia grazie ad un terrazzo fatto erigere davanti al Tempio di Apollo[50].

Apollo (*Sol*) giocava d'altronde un ruolo insostituibile nelle attività di fondazione e di consacrazione (dal momento che determinava l'orientamento del decumano e illuminava beneficamente la città dopo l'atto inaugurale).

È noto che Augusto scelse Apollo come divinità tutelare, tanto più che quest'ultimo

---

Tra le famiglie menzionate nelle iscrizioni di età protoimperiale si segnalano alcuni gentilizi latini (come quelli dei *Cornelii,* dei *Livii,* dei *Vibii,* degli *Octavii,* dei *Domitii,* degli *Hostilii,* degli *Avilii,* e dei *Lollii*), alcuni dei quali immigrati dal centro Italia, altri sicuramente trasferitisi in Piemonte dal Veneto, altri ancora risultato di adozioni da parte di *gentes* romane.

La prosopografia caratterizzata dall'assenza di veterani è stata considerata da alcuni come un indizio per interpretare la natura e la cronologia della fondazione augustea (la quale dovrebbe distaccarsi dal 'periodo bellico' di Augusto e posizionarsi in un momento successivo, quando si rese necessaria un'ingente assegnazione di terre ai soldati che avevano accompagnato il principe nelle sue imprese, cfr. DION. CASS. 53, 25). Tale osservazione è stata da altri ritenuta poco significativa in considerazione del fatto che degli oltre 300.000 cittadini che Augusto congedò, nelle epigrafi sin qui scoperte, pochissimi di loro fa menzione del proprio passato militare (cfr. PACI 1988, p. 113).

[46] In una recente revisione atta a ricostruire l'elenco delle colonie dedotte dai triumviri e da Ottaviano Augusto, E. Folcando elenca le fondazioni o le città probabilmente ristrutturate da Augusto: *Ariminium, Ateste, Augusta Praetoria, Augusta Taurinorum, Beneventum, Bononia, Brixia, Capua, Cremona, Iulia Dertona, Minturnae, Nola, Parma, Pisaurum, Placentia, Pola, Puteoli, Suessa, Sutrium, Teanum, Venafrum* e, forse, *Aquileia, Lucus Feroniae* e *Tergeste.* A questo elenco proporremmo di aggiungere plausibilmente *Augusta Bagiennorum* in Piemonte. E. FOLCANDO, 1996, pp. 75-76 e 105-106.
[47] AUG. *R.G.* 3.
[48] *Ibid*. 28.
[49] FEST. *de verb. sign.* 310.
[50] CARANDINI, 2008, pp. 89-94.

era intervenuto a suo favore durante la battaglia di Azio decretando la sconfitta di Antonio e di Cleopatra. Il racconto di Svetonio sulla nascita di Augusto, stilato sulla scia di un cliché stereotipato e consolidato sin dall'età di Alessandro Magno, non lascia dubbi: «La madre Azia, recatasi a mezzanotte ad una solenne cerimonia in onore di Apollo, fatta posare la lettiga dentro il tempio, si era addormentata mentre le altre matrone se ne andavano a casa. Un serpente, all'improvviso, le era strisciato addosso e poco dopo se ne era andato […]. Augusto che nacque nel decimo mese dopo questo fatto venne quindi stimato figlio di Apollo[51]». Dione Cassio lascia intendere che il prodigio suggestionò Caio Giulio Cesare al punto da indurlo ad adottare il giovane lasciandogli in eredità i propri beni e quelli dell'impero[52].

L'amore di Augusto per Apollo sembra essere un tema precoce: Svetonio ricorda come in una cena segreta tenutasi nel 39-38 a.C., Ottaviano si travestì da Apollo mentre gli altri convitati indossavano i costumi degli dei e delle dee dell'Olimpo. Così, considerato che in città il grano scarseggiava, il popolo si lamentò del fatto che «gli dei avevano mangiato tutto il grano e che Cesare era certamente Apollo, ma Apollo carnefice[53]».

Nel 28 a.C., in occasione di una sistematica ristrutturazione della *domus* acquistata sul Palatino resasi necessaria dopo la caduta sul caseggiato di un fulmine 'fatale', Ottaviano promosse la costruzione di un tempio dedicato ad Apollo[54] e fece disporre nella vicina biblioteca una statua *habitu ac statu Apollinis*.

L'interesse di Augusto per il cielo e l'ordine dei pianeti è d'altronde evidente nell'*horologium* monumentale fatto costruire in Campo Marzio sotto la direzione del matematico Facondio Novo[55]. La grande meridiana è stata ricostruita da E. Buchner negli anni Settanta[56]; il suo analemma occupava una piazza pavimentata in travertino con le strisce e le lettere in bronzo dorato disposte su un'area di 165 × 74 m. L'ombra proiettata sul selciato da un imponente obelisco egizio alto 100 piedi (opportunamente posizionato sul lato meridionale della meridiana) segnava l'ora del giorno; la stessa ombra, se osservata al mezzogiorno, indicava il segno zodiacale del mese in corso (i dodici segni erano, infatti, riprodotti a mosaico sull'analemma).

I recenti studi di B. Frischer dimostrano che l'obelisco della gigantesca meridiana era stato dedicato da Augusto al dio Sole (Apollo) mentre le simulazioni al computer chiariscono che il complesso era progettato in modo tale che il Sole, alla levata del solstizio invernale, si presentasse a chi proveniva da est (lungo la via Flaminia) tangente all'apice dell'obelisco ed in perfetto allineamento con la facciata dell'*Ara Pacis Augustae*.

---

[51] Svet. *Aug.* 94; Papini, 2013, pp. 219-224.
[52] Dio. Cas. 45, 1.
[53] Svet. *Aug.* 70: *Cena quoque eius secretior in fabulis fuit, quae vulgo* δώδεκαθεῶς *vacabatur; in qua deorum dearumque habitu discubuisse convivas et ipsum pro Apolline ornatum non Antoni modo epistulae singulorum nomina amarissime enumerantis exprobrant, sed et sine auctore notissimi versus*: […] *sexque deos vidit Mallia sexque deas, impia dum Phoebi Caesar mendacia ludit.*
[54] Svet. *Aug.* 29: *Templum Apollinis in ea parte Palatinae domus excitavit, quam fulmine ictam desiderari a deo haruspices pronuntiarant*. Inoltre Svet. *Aug.* 52: *Atqe etiam argenteas statuas olim sibi positas conflavit omnis exque iis aureas cortinas Apollini Palatino dedicavit.*
[55] Plin. *nat.* 37, 72; Moretti, 1947; Rossini, 2006. Portato a Roma dall'Egitto nel 10 a.C., l'obelisco era stato dedicato al Sole fra il 26 giugno del 10 e il 25 giugno del 9 a.C., come si ricava dall'indicazione della *tribunicia potestas XIV* di Augusto nell'iscrizione posta sulla base (*CIL* VI 702).
[56] Buchner, 1980-82, pp. 331-345.

Chi, lo stesso giorno dell'anno si fosse posizionato sul lato opposto dell'Ara (e dunque avesse puntato lo sguardo a ovest) avrebbe, per converso, visto l'ombra dello gnomone attraversare assialmente il grande portale aperto sulla facciata fino ad andare a toccare l'altare vero e proprio. Un fenomeno indubbiamente scenografico, che riconferma gli interessi astronomici del principe.

La volontà del complesso era probabilmente quella di sottolineare come un disegno provvidenziale avesse portato Augusto a far trionfare la pace: l'obelisco egizio rimandava idealmente alla vittoria sull'Egitto e su Marco Antonio, mentre la proiezione della sua ombra sull'altare della Pace celebrava il fatto compiuto[57].

### Misura strumentale dell'orientamento di *Augusta Taurinorum*

Per lo studio puntuale delle relazioni intercorrenti tra l'orientamento della città e il cielo antico è necessario disporre di un'accurata misurazione dell'azimut del cardo e del decumano.

Il valore angolare formato dall'asse del decumano di *Augusta Taurinorum* rispetto ai punti cardinali è accennato in alcune pubblicazioni scientifiche o indirettamente ricavabile da alcuni studi afferenti l'orientamento della centuriazione torinese.

Nello specifico, la centuriazione di *Augusta Taurinorum* è citata per la prima volta da P. Fraccaro (nello storico studio sulla centuriazione di Ivrea[58]) con una *pertica* orientata di 26° E S-E; tale misura è riproposta negli stessi termini da P. Lavedan e J. Hugueney nella *Histoire de l'urbanisme* edita a Parigi nel 1966 e viene giustificata in funzione di un presunto orientamento solstiziale invernale di *Augusta Taurinorum*. Il *cardo* di *Augusta Taurinorum* è quindi presentato con un'orientazione di 26° E S-E nel 1968 all'interno del celebre volume *Forma Urbana ed architettonica nella Torino barocca* edito da A. Cavallari Murat[59]; lo stesso valore è richiamato, quattro anni più tardi, anche da G. A. Mansuelli[60].

Nel contesto di più recenti ricerche archeoastronomiche, G. Magli ha computato un valore di 30° E S-E[61], misura successivamente rivalutata da A.C. Sparavigna in 25,8° E

---

[57] BUCHNER, 1980; REHAK, 2006; FRISCHER, FILLWALK, 2013.
[58] FRACCARO, 1941.
[59] CAPPA BAVA, 1968, p. 333.
[60] MANSUELLI, 1971.
[61] MAGLI, 2007.

S-E[62].

Di fronte ad una simile oscillazione di valori e per scongiurare che un errore di valutazione inficiasse il paradigma interpretativo, si è innanzitutto provveduto alla verifica strumentale dell'angolo formato dalla *pertica* di *Augusta Taurinorum*.

Come primo passo si sono importati all'interno del software AutoCAD i capisaldi geodetici pubblicati nel geoportale del Comune di Torino (visualizzatore SIT, Carta Archeologica) ottenendo un valore di 27,3° (angolo formato dal decumano con l'asse EW)[63].

Considerato che tale valore si discosta sensibilmente da quelli sino ad oggi pubblicati, si è reputato di dover procedere ad un'ulteriore verifica cambiando radicalmente metodo, ovvero appoggiandosi ad una moderna strumentazione GPS del tipo a doppia frequenza con metodologia WRS la cui precisione può stimarsi nell'ordine del centimetro[64].

Per effettuare la misurazione si è partiti dal presupposto che esista una sostanziale coincidenza tra di tracciato tra l'attuale via Garibaldi (già via Dora Grossa) e l'antico decumano romano.

Tale affermazione è confortata dalla planimetria pubblicata da C. Promis nel 1868[65] – poi integrata da A. D'Andrade[66] – nella quale i tratti di mura, delle strade e delle cloache romane sino a quel momento scoperti sono sovrapposti graficamente alla rete viaria moderna[67] (fig. 14).

Nel caso specifico, il basolato che corre lungo l'asse mediano di via Garibaldi/decumano appare quasi completamente documentato, dall'estremo est a quello

---

[62] SPARAVIGNA, 2012. Nel lavoro sopracitato, il valore di azimut del decumano dichiarato, ricavato utilizzando Google Earth, devia di quasi 2° da quello misurato con il GPS, probabilmente perché l'orientamento della strada è calcolato per mezzo della trigonometria piana su un'immagine della mappa e Google Earth non costituisce uno strumento di georeferenziazione scientifico (come d'altronde dichiarato da Google nelle note di rilascio). Nel preprint si propongono pertanto una serie di date (il 10 novembre, il 30 gennaio e il 21 dicembre) senza poter effettuare una scelta. Il 21 dicembre viene ricavato sostituendo la levata ortiva azimutale del Sole con quella dell'angolo orario: una soluzione che sembra poco plausibile (e che l'autrice stessa, d'altronde, introduce con qualche riserva), sia perché siffatta scelta vanificherebbe gli aspetti visuali dell'esperienza sacrale (i sacerdoti avrebbero dedotto il punto ortivo del Sole senza una misura in loco corrispondente in quel dato giorno e senza poterlo osservare nel cielo, ma basandosi su un conteggio di giorni a partire dall'equinozio di primavera), sia perché eluderebbe il rigido conservatorismo della disciplina etrusco-romana le cui radici affondano nel periodo regio e orientalizzante (almeno a partire dal VI sec a.C.), vale a dire prima della nascita della scienza ellenistica. A.C. Sparavigna non riconosce la festa augustea non potendo così addivenire ad una proposta sull'anno di fondazione per mezzo di un termine *post data* (nel nostro caso il 13 a.C.) e, non utilizzando la data giuliana, non può proporre un anno di fondazione specifico. L'uso del Sole vero computato nel presente studio sostituisce la data del 10 novembre con quelle dell'11, 12 e 13 novembre. Con l'uso del Sole vero, invece, la data del 30 gennaio perde il carattere di approssimazione per trasformarsi in un valore puntuale ripetuto. Anche la dedica di *Augusta Taurinorum* alla festa di Pax non è riconosciuta.
[63] Computo a cura di Nicola Pozzato.
[64] Rilevazione a cura di *Mejors and Monitoring* di Marco Ugolotti.
[65] PROMIS, 1869.
[66] MERCANDO, 1998, pp. 55-59.
[67] La documentazione sugli affioramenti del basolato nel settore occidentale del decumano provengono unicamente dalla bibliografia storica e non disponiamo, allo stato attuale, di dati archeologici georeferenziati. Gli appunti e i rilievi di A. D'Andrade e C. Promis, a dire il vero sufficientemente copiosi, sono da considerarsi del tutto affidabili perché si sono sempre mostrati pragmaticamente puntigliosi e accurati.

ovest, e presenta un orientamento coincidente con quello della strada attuale.

Anche nei rilievi archeologici e nelle foto scattate in occasione dei sondaggi condotti allo sbocco di via Garibaldi su piazza Castello dalla Soprintendenza Archeologica del Piemonte appare evidente l'allineamento tra la porta romana inglobata in Palazzo Madama e il basolato della via romana (fig. 15). Quest'ultima, a seguito dei rettifili delle facciate di età barocca, si trova spostata verso meridione di una misura stimabile tra gli 1,5 e i 2 m[68].

Per quanto concerne l'estremità occidentale del decumano, è bene ricordare che questo terminava in corrispondenza della Porta Segusina (detta nella 'bibliografia prevalente' Decumana). Oltre ai rilievi puntuali del basolato prossimali documentati da D'Andrade e Promis nella celebre china dei primi del Novecento (fig. 14), disponiamo del rilevamento dei resti della fondazione della porta romana orientale negli scantinati di via Garibaldi 39 curati nel 1897 dal collaboratore di A. D'Andrade, l'ingegner C. Bertea. La relazione autoptica riferisce che l'asse mediano della via si trovava un paio di metri a sud dell'asse di via Garibaldi[69]. Ne consegue una coincidenza dell'orientamento della via moderna con quello del decumano romano puntuale persino oltre le aspettative.

Anche le recenti ricerche archeologiche d'emergenza condotte presso il cosiddetto Quadrilatero romano confermano il quadro sin qui tracciato.

Particolarmente evocativa è, ad esempio, la fotografia scattata durante i lavori di risistemazione di via XX Settembre (fig. 16) in cui si vede con chiarezza l'asse dell'antico decumano minore correre lungo la linea mediata dell'arteria moderna

Gli estremi del segmento oggetto della misurazione sono stati fissati, pertanto, alle estremità di via Garibaldi (asse centrale della via all'intersezione con piazza Castello, asse centrale della via all'intersezione con piazza Statuto). Un'ulteriore misurazione di verifica è stata effettuata all'incrocio tra via Garibaldi con via Consolata (figg. 11-13).

Il valore ottenuto corrisponde ad un azimut di 117°40'46'' ovvero 117,6794°  117, 68° (vale a dire 27,68° E S-E) (figg. 7-10).

La ricalibrazione dell'orientamento della *pertica* torinese sul valore di 27,68° E S-E è indispensabile ai fini del discorso qui intrapreso e sembra stimolare ulteriori ricerche su quello della centuriazione cittadina (in genere l'orientamento del reticolo urbano e della centuriazione coincidono, ma l'esperienza dimostra che sono anche possibili deviazioni più o meno segnate). I valori angolari sono infatti fondamentali nell'ambito di qualunque discussione sul rapporto intercorrente tra la centuriazione di Ivrea e quella di Caselle e, più in generale negli studi sulla cronologia dell'*ager taurinense*.

---

[68] FILIPPI, 1982.
[69] BENDINELLI, 1929, p. 26; GRAZZI, 1981, p.75.

**Le misure astronomiche**

Per risalire al giorno di fondazione di una città romana secondo quanto enunciato dalla gromatica e della trattatistica antica occorre individuare in quali date l'angolo di levata del Sole corrisponda al valore misurato dell'angolo E S-E del decumano. Questo, in termini astronomici, equivale a ricavare il valore dell'azimut del Sole nascente al momento della fondazione di *Augusta Taurinorum*, vale a dire l'arco di orizzonte compreso tra l'Est e il punto in cui sorge il Sole o ampiezza ortiva.

Alla latitudine di Torino ($\lambda_{TO}$ = 45,07°) e ad una data altezza del Sole ($h_\odot$) la formula da utilizzare per l'azimut della sua levata è la seguente:

$$A = \frac{180}{\pi} \arccos\left(-\frac{\sin\delta}{\cos\lambda_{To}\cos(-h)} + \tan\lambda_{To}\tan(-h)\right) \quad (1)$$

da cui si ricava l'ampiezza ortiva $AO_\odot = A_\odot - 90°$.
La formula (1) dipende dalla declinazione[70] del Sole data da

$$\delta = \arcsin(\sin E \sin L) \quad (2)$$

dove *E* rappresenta l'obliquità media dell'eclittica (ovvero la sua inclinazione media rispetto all'equatore celeste) e *L* la longitudine eclittica geocentrica apparente (corretta per l'aberrazione) del Sole, entrambe calcolate secondo gli algoritmi che forniscono le correzioni per il Sole Vero[71]. Descrivendo il moto del Sole, tali quantità dipendono dal tempo, che in astronomia è computato per mezzo della Data Giuliana (*JD*), un sistema di datazione che conteggia il numero di giorni trascorsi a partire dal 4713 a.C.[72]. Essendo costituita da giorni civili di 24 ore, essa identifica univocamente l'istante in cui è avvenuto o avverrà un certo fenomeno astronomico indipendentemente dal calendario in uso (nel nostro caso riferita all'epoca standard JD2000).

Ciò detto, la formula (1) è stata implementata in linguaggio IDL (Interactive Data Language) per poter essere elaborata numericamente in funzione delle date giuliane che vanno dal 27 a.C. al 5 a.C., ovvero a partire dalla data di conferimento del titolo di

---

[70] Coordinata contata a partire dall'equatore celeste lungo il cerchio massimo passante per l'astro e la sua proiezione su di esso.
[71] LASKAR, 1986, p.149; BRETAGNON, SIMON, 1986; LATTANZI, PANNUNZIO, 2009, pp.13-24, 31 e 194. Si veda l'appendice A per la definizione e maggiori dettagli.
[72] La data giuliana è un sistema di datazione ideato da Giuseppe Scaligero nel 1583 d.C., in onore di suo padre, Giulio Cesare Scaligero, insigne matematico dell'epoca. I giorni civili dell'Anno Giuliano vengono contati progressivamente a partire dall'inizio del giorno (mezzogiorno) del primo di gennaio del 4713 a.C., o in tempi recenti dalle ore 12 del Tempo delle Effemeridi, accumulandoli per 7980 anni con giorni dell'Anno Giuliano fino al 4 ottobre 1582 e con giorni dell'Anno Gregoriano dal 15 ottobre del 1582 in poi (LATTANZI, PANNUNZIO, 2009, pag.62*)*.

Augusto ad Ottaviano (ritenuto dagli storici come un accettabile *post quem* considerata l'intitolazione di *Augusta Taurinorum*) e il 5 d.C. (una cronologia ritenuta tarda e posteriore alla fondazione nel dibattito relativo alla suddetta fondazione; Cfr. §1).

A rigore, oltre al moto vero, vi sono ancora diversi fattori che possono inficiare l'osservazione corretta dell'apparire del disco del Sole, in particolare: la rifrazione astronomica[73] e l'errore umano nella rilevazione del bordo solare (l'occhio umano riesce a percepire ad occhio nudo circa un arcominuto) a cui si deve aggiungere, nel caso di Torino, l'elevazione della collina che altera l'orizzonte astronomico. Va tenuto conto anche che le variazioni giornaliere del tempo atmosferico o altre cause influenzano e determinano gli orari del sorgere del sole.

Ora, se si assume che il sorgere del Sole abbia luogo nel momento in cui il suo bordo superiore appare all'orizzonte, l'istante dell'alba dovrà manifestarsi con un po' di anticipo, tenendo conto della rifrazione.

Al suolo, i valori angolari che si devono sottrarre sono di 34' per la rifrazione e di 16,2' per il raggio del Sole (equivalente a 0,27°, calcolato dalle formule per il Sole Vero alla latitudine di Torino nei giorni giuliani considerati), sicché in totale se ne ricava un'altezza al suolo, considerata la sola rifrazione, pari a $h_{rif}$ = 50' ~ 0,83°.

Un secondo problema nasce dal fatto che a Torino il sole sorge dietro una serie di colline il cui profilo, molto variabile, non consente un'esatta valutazione dell'altezza da considerare per il calcolo del tempo dell'alba. A ciò si deve aggiungere che il popolamento arboreo di età recente e subrecente ha alterato leggermente le prospettive visuali, in positivo e in negativo. D'altronde, è chiaro, che non ci è possibile ad oggi ricostruire la copertura arborea della collina torinese nell'antichità.

Esplorando cartograficamente il prolungamento del decumano verso il punto di levata, se ne evince che il punto dietro al quale sorge il Sole è un picco all'altezza circa 490 m s.l.m., posto all'incirca alla stessa altitudine della cittadina di Pino Torinese (501 m s.l.m.) che è anche il punto geografico più estremo dell'asse visuale, perché qui ha fine la collina torinese (fig. 17). Pertanto si può desumere dalla mappa (fig.17) una distanza tra il punto ortivo di levata del Sole all'alba e l'*umbilicus* (altezza sul livello del mare circa 238 m) pari a circa 8300 m (considerando un incertezza di ±200 m).

A titolo di curiosità, può essere interessante notare che, il prolungamento della linea del decumano torinese lambisce il versante collinare proprio alla base dell'Osservatorio Astrofisico di Torino (un caso della sorte veramente curioso, tanto più che Palazzo Madama fu la sua precedente sede).

Ricapitolando, l'angolo da cui è visibile il picco a 490 m s.l.m. dall'*umbilicus* è di 1,7°, ma a tale valore va sottratto un angolo di rifrazione pari a 0,34° (stimata con valori standard di pressione atmosferica e temperatura per Torino) e un ulteriore angolo di 0,27° per il raggio angolare del Sole. In tal modo l'altezza effettiva del Sole da considerare diventa 1,09° = (1,7° - 0,34° - 0,27°).

---

[73] Come è noto, i raggi luminosi emessi da un corpo celeste, prima di raggiungere l'osservatore, passano attraverso l'atmosfera terrestre e vengono rifratti. Ciò è dovuto alla densità dell'atmosfera che aumenta verso la superficie terrestre, per cui il raggio luminoso devia dal percorso rettilineo che percorrerebbe in assenza di atmosfera e alza, in sostanza, l'astro sopra l'orizzonte (LATTANZI, PANNUNZIO, 2009, pp. 95-96).

I dati generati con il programma IDL (riportati in appendice B) sono stati poi selezionati strettamente intorno alla direzione misurata del decumano entro un intervallo pari a circa ±0,07°, assunto come valore angolare parallattico della larghezza del decumano (circa 10 metri) rispetto alla lunghezza dell'ostacolo (circa 8300 metri); di seguito, sfruttando la libreria astronomica di IDL nel programma elaborato, è stata effettuata un'inversione numerica della fomula (1) da data giuliana a giorno civile, per gli anni summenzionati, fissando come dato di azimut il valore 27, 68° E S-E (valore da assumere quale risultato finale della misura effettuata dai gromatici, includente tutti i possibili errori). Inoltre l'altezza effettiva di 1,09° è stata variata applicando un passo progressivo di 0,01° come accuratezza intrinseca del programma IDL (con passi più piccoli i risultati non modificano le prime due cifre significative dopo la virgola) da 1,06° fino a 1,18°. In questo modo si è voluto considerare anche un margine di errore un po' più ampio che includa l'angolo di 3' corrispondente alla percezione della prima apparizione del bordo del Sole sul profilo collinare e una occasionale scarsa visibilità atmosferica. Su quest'ultimo aspetto vale forse la pena notare che l'elevazione considerata rappresenta il punto più alto del profilo collinare tra Superga e l'Eremo; quindi, probabilmente, era anche il punto E S-E in cui, in condizioni metereologiche avverse, il Sole risultava meglio individuabile al momento della misurazione della sua levata.

Si evidenzia che l'uso congiunto del Sole vero e dell'inversione della formula (1) in funzione della data giuliana permette di definire con maggiore precisione gli anni più probabili per la fondazione della città, oltre che il giorno, quest'ultimo desumibile anche con la sola applicazione di un Sole Medio attraverso le classiche formule di trigonometria sferica e i coefficienti opportuni per stimare l'obliquità dell'eclittica.

Questo determina l'unicità del nostro studio perché, essendo l'intervallo di date di fondazione possibili imposto dal contesto storico piuttosto ristretto (l'età augustea dal 27 a.C. al 5 a.C.), è stato possibile effettuare una scrematura scientifica delle date più probabili evidenziando dati inequivocabili che si accordano con i contesti geografici, epigrafici, storici e archeologici disponibili.

**L'interpretazione dei dati. Il 30 gennaio del 9 a.C.**

Come si evince dalle tabelle pubblicate in appendice B, più date si propongono come possibili candidate per la fondazione e l'inaugurazione di *Augusta Taurinorum*. In particolare, ogni anno si verificano almeno due date compatibili con l'orientamento del *decumano* cittadino, delle quali una in autunno (12 e 13 novembre) e una in inverno (30 gennaio). Ciò deriva dal moto stesso del Sole che percorre archi gradatamente sempre più alti a partire dal solstizio invernale e procede in senso inverso nei sei mesi che seguono al solstizio estivo.

Inoltre, l'aver effettuato il computo archeoastronomico considerando il moto del Sole vero alla latitudine di Torino ha come conseguenza una certa variabilità delle date ricavate di anno in anno. Tale approccio rende tuttavia possibile l'individuazione dell'anno corrispondente all'altezza del Sole nel giorno di fondazione della città sulla base delle reali condizioni locali di misura e in armonia con l'azimut del decumano che ci è stato possibile misurare.

Una volta ottenuti dei dati scientifici consistenti, per definire quale data – invernale o autunnale – sia da considerarsi valida, ci si è necessariamente affidati alla critica storica e archeologica, vale a dire ad un processo di carattere interpretativo.

Adottando tale metodo, salta all'occhio come la data del 30 gennaio sia straordinariamente pertinente al periodo augusteo dal momento che coincide con una delle più importanti festività romane[74].

Nella Tabella β pubblicata in appendice sono riassunti i principali eventi ricordati dalle fonti antiche che si collegano strettamente alla vita del principe, sia per quanto concerne la sfera privata, sia per quanto concerne quella pubblica. Di tali date, un buon numero deve essere necessariamente scartato in quanto relativo ad eventi infausti (quali le sconfitte militari o il decesso di congiunti).

La data del 30 gennaio coincide invece con la celebrazione della festa dell'*Ara Pacis Augustae*, una ricorrenza istituita dal Senato in onore di Augusto nel 13 a.C. e celebrata a partire dal 30 gennaio dal 9 a.C. in Campo Marzio dai magistrati, dai sacerdoti di Roma e dalle Vestali[75]. La data è precisata dai Fasti Prenestini ove si legge *Ad III Kal. Feb.— Feriae ex S.C. quo[d eo] die ara Pacis Augusta[e in campo] Martio dedicata [e]st Druso et Crispino c[oss.].*

Dal punto di vista storico, un collegamento tra *Augusta Taurinorum* e la festività di *Pax* è piuttosto significativo: l'istituzione della festività medesima fu infatti la conseguenza di un lungo periodo di guerre che si conclusero con la conquista delle Alpi per mezzo di operazioni militari che ebbero come teatro anche la valle di Susa e i territori contermini (si veda a tal proposito l'iscrizione dell'arco di Susa[76]).

Ora, se accettiamo come valida la data del 30 gennaio (data che, come si vedrà, è sostenibile grazie a diversi argomenti di carattere storico e contestuale), essa viene automaticamente a costituire un termine *post quem*, dal momento che l'inaugurazione di *Augusta Taurinorum* non poteva precedere l'istituzione della festa medesima che fu decretata nel 13 a.C.

Se passiamo poi ad una lettura dei dati archeoastronomici ricavati dallo studio del cielo antico, le date del 9, del 5 e dell'1 a.C. (fig. 18) risultano le migliori candidate per il rito dell'*inauguratio* poiché, in loro coincidenza, l'orientamento dell'asse decumano/via Garibaldi manifesta il minimo scarto con l'azimut del sole alla sua levata. Di queste però è la data del 30 gennaio del 9 a.C. che manifesta il maggior grado di pertinenza (in relazione anche alle altezze effettive), sicché dovrà essere considerata la più attendibile.

---

[75] AVG. *R.G.*, XII: *Cu]m ex H[is[]ania Gal[liaque, rebu]s in iis provincis prosp[e]re [gest]i[s], R[omam redi] Ti. Nerone P. Qui[ntilio c]o[n]s[ulibu]s, ~ aram [Pacis A]u[g]ust[ae senatus pro]redi[t]u meo consa[c]randam [censuit] ad campam [Martium, in qua ma]gistratus et sac[er]dotes [et v]irgines V[est]a[les ann]iversarium sacrific]ium facer[e decrevit.].* Cfr. CENTANNI, CIANI, 2017.
[76] CRESCI MARRONE, 1994, pp. 185-196; FOGLIATO, 1992.

Se le cose stanno così, ne conseguirebbe un certo grado di sincronia tra la data di deduzione di *Augusta Taurinorum*, la celebrazione della festa di *Pax* e – in ultima analisi – l'inaugurazione dell'arco onorario della vicina *Segusio* che avvenne nel 9-8 a.C.

Si tratta di una situazione eccezionale, derivata da una serie di concomitanze fortunate, quali la possibilità di fondare la colonia in un territorio relativamente libero da preesistenze (che permise di scegliere l'orientamento del decumano senza troppi vincoli), l'interesse del principe nei confronti della *disciplina* etrusco-romana e della tradizione avìta di fondazione e di inaugurazione, nonché l'istituzione da parte di Roma – negli stessi anni – di una nuova festività di carattere squisitamente politico.

## Il tema astrologico di *Augusta Taurinorum*: Augusto, il Sole e il Capricorno

L'ipotesi di un orientamento *ex caelo* di *Augusta Taurinorum* è stata stimolata da uno studio condotto sulla vicina *Augusta Praetoria Salassorum*, una colonia fondata nel 25 a.C. successivamente alle campagne militari del console Aulo Terenzio Varrone Murena nel territorio dei Salassi, all'imbocco dei passi alpini della *Alpis Graia* (Piccolo S. Bernardo) e dell'*Alpis Poenina* (Gran S. Bernardo).

Recenti scavi condotti in corrispondenza delle fondazioni della cosiddetta Torre dei Balivi hanno permesso di portare alla luce un insospettato rilievo raffigurante il segno del Capricorno affiancato da un aratro (ne sono ancora riconoscibili il manico e forse il vomere) e da due falli apotropaici, simboli interpretati dagli scopritori come dei 'sigilli' benaugurali contestuali alla deduzione della colonia[77] (fig. 20). Il ritrovamento ha stimolato uno studio archeoastronomico che ha dimostrato che *Augusta Praetoria* fu orientata rispetto al corso del sole (considerando e computando, tra l'altro, l'elevazione delle montagne circostanti) in modo da ottenere l'allineamento del *cardo maximus* con il sole il giorno del solstizio invernale: un giorno dell'anno particolarmente caro ad Augusto e frequentemente celebrato nelle gemme e nei monumenti celebrativi (cfr. §. 8)[78].

Non lontano da Torino, una terza colonia augustea (*Augusta Bagiennorum*) presenta un decumano orientato verso la stessa porzione di cielo (anche se lo studio è parziale, poiché l'interro di molte strutture romane non permette la misurazione puntuale degli

---

[77] BERTARIONE, 2012; EAD., 2013.
[78] BERTARIONE, G. MAGLI, 2014.

orientamenti stradali)[79].

A questi esempi si aggiunge, infine, un recente studio che documenta un orientamento dell'*Ara Pacis Augustae* in corrispondenza dell'azimuth di levata del sole al 21 aprile, genetliaco della stessa Roma[80].

Nel caso di *Augusta Taurinorum,* la coincidenza del 30 gennaio con un'importante festa augustea e la levata eliaca del segno del Capricorno, a cui seguiva il sorgere del Sole, rappresentano un ulteriore indizio di intenzionalità e rafforzano le osservazioni precedenti.

Ricordiamo che per levata eliaca si intende l'apparizione di un astro o di una costellazione subito prima del sorgere del Sole. L'astro o la costellazione sono unicamente visibili in tale circostanza perché, trovandosi nel corso della giornata nella porzione di cielo sopra l'orizzonte, risultano di fatto invisibili a causa della luminosità del Sole.

Considerata la precessione degli equinozi, a fine gennaio ciò accadeva per l'asterismo del Capricorno, che essendo congiunto al Sole nei trenta giorni precedenti, ne seguiva il corso rendendosi invisibile per tale lasso di tempo, per poi diventare appena visibile prima dell'alba, quando in congiunzione al Sole risultava essere l'adiacente costellazione dell'Acquario (ovvero a fine gennaio). Ciò è dovuto al moto apparente del Sole lungo l'eclittica che nell'arco dei 12 mesi va a coprire le ben note dodici costellazioni.

Anticamente, la levata eliaca di un astro era di grande importanza pratica e fungeva da orologio notturno; nel nostro caso, probabilmente, indicava il periodo di tempo antecedente al sorgere del Sole e, probabilmente, il momento nel quale era necessario apprestarsi ad effettuare le misure necessarie per tracciare il decumano.

Lo studio del cielo e delle congiunzioni astrali era piuttosto diffuso tra gli aristocratici dell'antichità, come d'altronde testimoniano gli scritti di Varrone[81] (che realizzò nella sua villa di Cassino una voliera dotata di un padiglione che gli permetteva di scrutare e studiare il cielo) e il trattato didattico sulle Arti Liberali compilato nel V sec d.C. da Marziano Capella (che ebbe ampia diffusione nel Medioevo)[82].

Lo studio dell'astrologia/astronomia assunse in Augusto aspetti peculiari di cui siamo informati grazie ad una serie di ricerche stimolate dal ricorrere del bimillenario della sua morte (nel 2014) che hanno avuto il merito di arricchire le nostre conoscenze sul programma politico e propagandistico dell'imperatore, oltre che sulle tendenze personali e gli interessi culturali.

Secondo quanto riferito da Svetonio, mentre nel 45-44 a.C. risiedeva ad Apollonia

---

[79] BARALE, CODEBÒ, DE SANTIS, 2001, pp. 489-502. La deviazione tra l'angolo misurato (pari a 130° 29') e quello previsto al solstizio d'inverno (127° 16') dichiarato dai ricercatori come data di fondazione della città è un po' troppo elevato. Tali misure – pur riproponendo il tema zodiacale in Capricorno tipico delle fondazioni augustee – dovrebbero forse essere soggette a verifiche più approfondite per valutare se l'orientamento non coincida piuttosto con una delle molte festività augustee che si concentrano nel mese di gennaio.
[80] VANCE TIEDE, 2016, pp. 267-274.
[81] VARRO. *de re rust.* 3, 5.
[82] MART. CAP. *de nupt. Phil. et Mer.*

(oggi in Albania) attendendo agli studi di grammatica e retorica, il giovane ebbe l'occasione di consultare l'astrologo *Theogenes* all'interno del suo osservatorio; in tale occasione, gli auspici furono talmente favorevoli da indurre l'astrologo a prostrarsi a terra[83]. L'esperienza dovette suggestionare profondamente il futuro principe che, secondo quanto riferito dallo storico romano, acquisì fede nel proprio destino e «divulgò quell'oroscopo facendo una moneta d'argento con il segno del Capricorno, sotto il quale era nato».

In effetti, soprattutto dopo la battaglia di Azio del 31 a.C., il tema zodiacale del Capricorno apparirà sempre più frequentemente nell'iconografia ufficiale: così ad esempio entro una sfera sospesa sulla testa del principe nelle celebre Gemma Augustea del *Kunsthistorisches Museum* di Vienna[84] (fig.21), nel cammeo in onice dell'*Art Institute of Chicago* opera di Dioscuride[85] (fig. 23), nella gemma in sardonica conservata nella collezione dell'Hermitage (in associazione ad un delfino, ad un altare e ad un tridente, in ricordo della vittoria di Azio, fig. 24[86]), negli aurei e nei denarii coniati nel 28 e nel 27 a.C.[87] (fig. 22), oltre che in svariati altri monumenti pubblici e privati[88]. Anche nell'ambito dell'architettura e della scultura il tema si ritrova, ad esempio, nel frontoncino di una tomba conservato nel Museo di Colonia[89], in un'antefissa augustea in terracotta rinvenuta ad Albano Laziale nel 1884[90] e in un frammento architettonico inserito nel tempio del foro di Ostia.

Augusto nacque il nono giorno prima delle calende di ottobre (23 settembre) poco prima del sorgere del Sole: *Natus est Augustus M. Tullio Cicerone C. Antonio consulibus VIIII Kal. Octob. paulo ante solis exortum, regione Palati ad Capita bubula*[91].

Anche l'astrologo Manilio nel «Poema degli astri» conferma che il segno zodiacale del principe era la Bilancia, ovvero che egli era nato *sub pondere librae*[92].

---

[83] SVET. *Aug.* 94: *In secessu Apolloniae Theogenis mathematici pergulam comite Agrippa ascenderat; cum Agrippae, qui prior consulebat, magna et paene incredibilia praedicerentur, reticere ipse genituram suam nec velle edere perseveravat, metu ac pudore ne minor inveniretur. Qua tamen post multas adhortationes vix et cunctanter edita, exilivit Theogenes adoravitque eum. Tantam mox fiduciam fati Augustus habuit, ut thema suum vulgaverit nummumque argenteum nota sideris Capricorni, quo natus est, percusserit.*

[84] BORBONE, 2013, p. 90.

[85] GROSS DIAZ, 2014, l'autrice propone che vi sia un esplicito riferimento all'avventura in Egitto degli dei olimpici i quali, minacciati dai Titani e da Tifone, furono salvati da Dioniso che si trasformò in uno strano pesce-capricorno (https://publications.artic.edu/roman/reader/romanart/section/522/end).

[86] NEVEROV, 1971, pp. 88-89.

[87] BORBONE, 2013, p. 91.

[88] *Ivi.,* pp. 88-90.

[89] Il Capricorno, d'altronde, divenne il contrassegno delle legioni *II Augusta* e *XIV Gemina Martia Victrix*.

[90] ROGER, 2013, p. 148.

[91] SVET. *Aug.* 5. È interessante notare che un editto del proconsole Paolo Fabio Massimo stabilì che il primo giorno del calendario romano provinciale fosse il giorno di nascita di Augusto (*IX kal. Oct*), vale a dire il 23 settembre, ridenominato in suo onore *Kaisar* (*OGIS* 458, 50-52). Sulla data di nascita di Gaio Ottavio cfr. P. Y. L. WARNE, 2017.

[92] Occupandosi del regno di Tiberio, lo stesso Manilio certifica la fine della giurisdizione del Capricorno a favore della Bilancia, segno sotto il quale era posto Tiberio: «Quando le chele dell'autunno stanno nascendo, benedetto è colui nato sotto il segno della Bilancia. Come giudice egli saprà giudicare con la bilancia e imporrà il peso della sua autorità sopra il mondo e legifererà. Le città e i popoli tremeranno al suo cospetto e saranno governati

Per spiegare tale apparente contradizione (vale a dire l'esaltazione publicistica del segno del Capricorno nonostante che il tema natale del principe fosse in Bilancia) è bene considerare che gli astrologi distinguevano tra il 'giorno del concepimento' e quello 'natale'[93]. Augusto era stato concepito a dicembre quando il sole era ospitato dal segno del Capricorno (il cosiddetto 'punto della fortuna' nella compilazione dell'oroscopo)[94] e forse anche la Luna si trovava nel medesimo segno nel giorno della sua nascita. Sicuro del fatto che gli astri fossero depositari del destino umano, Augusto riconosceva la data del suo concepimento come fortunato (in quanto coincidente con il giorno più breve dell'anno che annunciava l'inizio di un nuovo corso e, simbolicamente, di una nuova età). Il suo giorno natale vero e proprio era però il 23 settembre (sotto il segno della Bilancia, simbolo della Giustizia), cosa che si reputava faborisse il ruolo giocato del principe nel processo di ricomposizione sociale seguita al travagliato periodo delle guerre civili. Tutto ciò spiega il riproporsi di questi segni per tutto il corso della vita dell'imperatore.

Nell'ambito del discorso sin qui intrapreso, è significativo il fatto che il tema zodiacale del Capricorno sia implicitamente richiamato nella data di fondazione di *Augusta Taurinorum.* Dal punto di vista storico e archeologico, tale scoperta offre un dato incontrovertibile: vale a dire la riconferma (appoggiata dalla nomenclatura della città romana) che la fondazione dell'impianto urbano romano che ci è pervenuto (con un orientamento di 27,68° E S-E) ebbe effettivamente luogo sotto il principato di Augusto.

### Il quadro storico: una cronologia "bassa"

Abbiamo già accennato all'importanza del 30 gennaio nell'impianto propagandistico e celebrativo voluto da Augusto e al fatto che tale festività - concepita dal principe nel 13 a.C. -, fu celebrata annualmente a Roma e nelle province a partire dal 9 a.C. (vale a dire quattro anni più tardi).

La scelta di consacrare a *Pax* la neonata colonia taurinense si armonizza perfettamente con il contesto storico del periodo in cui essa venne fondata e assume un significato pregnante soprattutto se inquadrata nel contesto degli eventi politici che interessarono la regione alpina nel periodo tardo-repubblicano e primo imperiale.

È bene premettere che le Alpi si mantennero per lungo tempo alla periferia dell'area di controllo territoriale romano, sicché all'alba del principato augusteo le tribù indigene insediate nelle Alpi Cozie e Pennine erano ancora indipendenti pur essendo, di fatto, strette su entrambi i versanti da Roma[95].

---

solo da lui, e dopo la sua dipartita, la giurisdizione sui cieli gli spetterà» (MAN. *astronomica* 4).
[93] SVET *Aug.* 94: *Tantam mox fiduciam fati Augustus habuit, ut thema suum vulgaverit nummunque argentum nota sideris Capricorni, quo natus est, percusserit.*
[94] SCHÜTZ, 1991) pp. 55-67; PAPINI, 2013\*\*, p. 251.
[95] Strabone ricorda inoltre che nel 43 a.C. Decimo Giunio Bruto Albino, in fuga da Modena e intenzionato a

La scelta strategica avvenuta sotto il principato di Augusto di portare a compimento la conquista dei passi alpini[96] andando al di là delle già consolidate acquisizioni nelle Alpi Marittime[97] fu, in un certo senso, la diretta conseguenza delle campagne di conquista condotte da Giulio Cesare in Gallia, e si rese indispensabile per garantire il transito delle merci, dei viaggiatori e degli eserciti tra le province ubicate al di qua e al di là delle Alpi (di fatto, tra la Gallia Transalpina e la Gallia Cisalpina).

Le operazioni iniziarono nel 27 a.C. quando Augusto richiese ed ottenne dal Senato di Roma il controllo diretto delle province di Spagna (Terraconense e Lusitania) nonché quello della Gallia Narbonense, della Lugdunense, dell'Aquitania e della Belgica, presso le quali furono trasferite ingenti forze militari[98].

Nel 26 a.C., Aulo Terenzio Varrone Murena, su ordine di Augusto, penetrò in profondità nella terra dei Salassi in Valle d'Aosta facendo prigionieri 36.000 uomini[99], parte dei quali fu venduta sul mercato degli schiavi ad Ivrea. Tale campagna consentì la fondazione della colonia di *Augusta Praetoria* nella pianura alluvionale in cui convergevano le vie dirette ai passi dell'*Alpis Graia* e dell'*Alpis Poenina.* La città venne ad ospitare 3000 coloni romani (tra i quali si dovettero annoverare dei veterani scelti, provenienti dalle *cohortes praetoriae*[100]) oltre a un certo numero di indigeni che furono censiti e accolti nella nuova comunità.

Nel 20 a.C. Marco Vipsanio Agrippa risiedette a *Lugdunum*/Lione per sovrintendere alla costruzione di una nuova rete stradale ben ramificata diretta verso l'Aquitania, il Reno, la Belgica settentrionale e le foci del Rodano[101].

Tra il 16 e il 13 a.C. lo stesso Augusto soggiornò a Lione da cui ebbe la possibilità di riordinare con cura l'amministrazione dei territori gallici e analizzare con attenzione la situazione della fascia occidentale delle Alpi italiane.

---

congiungersi con le truppe di *Lucius Munatio Planco* che era accampato in Gallia fu costretto ad attraversare il territorio dei Salassi e venne costretto a pagare come pedaggio una dracma per ogni soldato del seguito (STR. 4, 6). Nel 35 a.C. Ottaviano Augusto dovette rinunciare ad imbarcarsi per l'Africa a causa di una insurrezione a cui parteciparono anche i Salassi nel 35 a.C. (DION. CASS. 49, 34). Una seconda rivolta (34 a.C.) fu repressa dal generale Valerio Massalla (*Ivi.*, 38).

[96] VOTA, 2000, pp. 11-46.
[97] Sia come effetto degli eventi della Seconda guerra punica, sia per la necessità di garantirsi un passaggio per via di terra in direzione della Spagna.
[98] DION. CASS. 53, 12. Già nel 22 a.C. la Provincia della Gallia Narbonese veniva restituita al Senato in quanto pacificata (LIV. 4, 1).
[99] STR. 4, 6; Cassio Dione aggiunge che in un primo momento Terenzio Varrone chiese unicamente un riscatto ma poi inviò i soldati romani presso i vari villaggi e fece arrestare gli uomini in età da armi vendendoli come schiavi con il vincolo di non essere liberati prima di vent'anni (DION. CASS. 53, 25).
[100] Le *cohortes praetoriae* erano spesso costituite da veterani scelti che continuavano a militare nelle legioni con alcuni privilegi di stipendio e trattamento, e molte di esse rimasero al servizio di Ottaviano e Marco Antonio dando luogo ad unità che, in qualche caso, raggiunsero l'esuberante numero di 200 unità (cfr. APP. 4, 115; VAN ROYEN, 1973, pp. 65-69). L'accoglienza di indigeni nel corpo civico è testimoniata dalla celebre iscrizione scoperta ad Aosta posta originariamente alla base di una statua dedicata ad Augusto dai *Salassi incolae qui initio se in coloniam contulerunt* (*Inscr. It.* XI 1. 6).
[101] STR. 4, 611.

Lo stesso anno, siamo al corrente di una serie di operazioni militari nelle Alpi affidate prevalentemente ai figliastri Tiberio e Druso che, partendo dalle basi militari sul Lago Lemano e da Aquileia, misero in atto una manovra a tenaglia che condusse alla progressiva sottomissione di 46 tribù alpine e che si concluse presso le sorgenti del Danubio, in Vindelicia. La vittoria fu celebrata (tra il 7 e il 6 a.C.) nel celebre *Tropaeum Alpium* a La Turbie, in Costa Azzurra.

Nel 13 a.C. le azioni militari si estesero dalle Alpi Marittime verso nord[102] sicché, il regolo celtico Cozio, posto a capo di una coalizione di tribù montane (di fatto ormai accerchiate), dopo un iniziale tentativo di guerriglia (a cui però accenna solo Ammiano Marcellino) strinse un patto di amicizia con Augusto, stipulando un *foedus* nel 13 a.C. che dava vita ad una prefettura romana. Quest'ultima, per riconoscenza, fu affidata al vecchio re che, adottato da Augusto ed entrato a far parte dell'ordine equestre, assunse il titolo di *praefectus*.

Lo stesso anno Augusto, di ritorno dalla Gallia e dalla Spagna pacificate, istituì la festività in onore di *Pax* avviando la costruzione dell'*Ara Pacis Augustae* nel Campo Marzio grazie a un decreto del Senato.

Il 30 gennaio del 9 a.C. venne celebrata per la prima volta la Festa della Pace presso l'*Ara Pacis* alla presenza del principe, della famiglia imperiale, dei sacerdoti, dei magistrati, delle vestali e del popolo di Roma.

Significativamente, nell'anno 9-8 a.C., fu inaugurato l'arco onorario di Susa in cui si celebra l'amicizia delle tribù celtiche della valle con Roma[103] e, poco dopo (attorno al 7 a.C.), si procedette alla divisione dell'Italia unificata nelle canoniche XI *Regiones* augustee, ascrivendo quella occidentale a nord del Po alla *XI Transpadana*.

Quello in questione può dunque, a ragione, considerarsi un periodo particolarmente denso di iniziative che coinvolsero, direttamente e indirettamente, la regione ai piedi delle Alpi. Così, tra il 20 e il 5 a.C., si procedette all'organizzazione di un sistema di dogana ai confini con la Gallia (a ovest di Torino presso *Ad Fines*, in reg. Drubiaglio di Avigliana) comportante l'esazione di un pedaggio il cui valore coincideva con il 2,5% di quello delle merci in transito (*Quadragesima Galliarum*)[104].

In ambito astronomico, è degno di nota che nell' 8 a.C. fu condotta la revisione del calendario solare elaborato dall'astronomo greco Sosigene di Alessandria al tempo di Giulio Cesare, interpolando un ulteriore mese bisestile per correggere l'errore compiuto involontariamente dai sacerdoti[105]. Il mese di *Sestile* venne quindi ribattezzato *Augustus* in onore dell'imperatore tramite la *lex Pacuvia de mense Augusto*[106].

Accettando che l'inaugurazione di *Augusta Taurinorum* sia stata effettuata con un orientamento del decumano tale da assecondare la festività del 30 gennaio, se ne

---

[102] DION. CASS. 54, 24.
[103] CARANZANO, 2016, pp. 13-46.
[104] J. FRANCE, 2001, p. 498.
[105] POLVERINI, 2016, pp. 95-144.
[106] SVET. *Iul.* 76, 1; DIO 44, 5; CENS. 22, 16; MACR. 1, 12c.

traggono alcune conseguenze di ordine storico particolarmente significative.

Innanzitutto si deve escludere una fondazione della colonia prima dell'istituzione della festa medesima nel 13 a.C.; con ogni verosimiglianza è anche poco probabile che la deduzione sia avvenuta antecedentemente alla sua prima celebrazione occorsa a Roma nel 9 a.C.

Come è stato accennato, l'interpretazione puntuale dei dati archeoastronomici offre come più probabili candidate per la deduzione della colonia le date del 9, del 5 e dell'1 a.C.; di queste, la data del 30 gennaio del 9 a.C., è quella che meglio si confà all'orientamento del decumano (Tavv. α,β) . La cronologia è perfettamente compatibile con i dati epigrafici, archeologici e letterari noti, e si armonizza perfettamente con l'opinione espressa ormai della prevalenza della critica storica, orientata a situare la fondazione della colonia taurinense nella metà del secondo decennio a.C.[107], probabilmente in concomitanza con le assegnazioni effettuate a favore di reparti di veterani congedati da Augusto (cosa che non esclude l'assegnazione di terre e proprietà anche a una certa percentuale di civili[108] come attestato ad *Augusta Praetoria* dove, nel 23 a.C., i Salassi «qui initio se in coloniam contulerunt» dedicarono una statua all'imperatore)[109].

Ubicata in un territorio attraversato da un'importante arteria commerciale e militare, posta al centro di una pianura fertile proprio all'imbocco delle valle di Susa (dove, come si è accennato, veniva esatta la *Quadragesima Galliarum*), dedotta al completamento delle guerre alpine (25-14 a.C.) e per tale motivo consacrata alla Pace, *Augusta Taurinorum* potrebbe insomma giustificarsi in qualità di tappa logistica lungo la strada diretta verso le Gallie venendo a costituire, al contempo, una risorsa economica per i coloni che vi si insediarono, i quali beneficiarono di un territorio fertile drenato dal corso del Po e dalla Dora Riparia.

D'altronde, non sfugge come la cronologia delle mura urbiche piuttosto alta (oggi sappiamo che la cinta e le porte di *Augusta Taurinorum* furono edificate tra l'età di Tiberio e quella dei Flavi, cfr. figg. 3-4) sia poco conciliabile con una colonia dedotta nel periodo turbolento della conquista della Gallia e delle Alpi. L'aspetto ricercato e in un certo senso decorativo delle porte e delle mura di *Augusta Taurinorum* si fa meglio comprensibile se inteso come frutto di un progetto stilato in tempo di pace finalizzato a una migliore definizione urbana della colonia e a regolare il suo rapporto con il territorio che la circondava[110]. Diversa sembra invece la situazione ad *Augusta Prateoria* che, dotata sin dall'inizio di torri capaci di ospitare balliste, di porte monumentali sovradimensionate e di un *agger* difensivo, manifesta un impegno costruttivo e fortificatorio[111].

---

[107] Si veda in particolare l'importante contributo di G. Masci che, partendo dalla documentazione storica, giunge a conclusioni coincidenti con quanto si evince dallo studio architettonico e archeoastronomico: MASCI, 2012, pp. 63-78. Sulla stessa linea di pensiero cfr. VOTA, 2000, pp. 11-46; CRESCI MARRONE, 1997., pp. 137-141; PACI, 1998, pp. 106-131; CULASSO GASTALDI, 1988., pp. 219-229.

[108] Come d'altronde sembra implicito nella presenza del nome dei Taurini nella denominazione ufficiale della città romana. Cfr. CULASO GASTALDI, 1998, p. 221.

[109] *Insc. It*. XI. 1. 6. MOLLO MEZZENA, 1981, pp. 139-141.

[110] CARANZANO, 2012, pp. 125-130.

**Conclusione ed "auspici"**

Come si è dimostrato, la data del 30 gennaio di un anno successivo al 13 a.C. (con ogni probabilità il 9 a.C.) costituisce un giorno credibile per l'inaugurazione della città, che fu fondata con la levata eliaca del segno del Capricorno e sotto l'auspicio della Pace.

Si reputa che Augusto abbia scelto la tale data per celebrare la Festa di *Pax* in considerazione che, il 30 gennio del 9 a.C., cadeva il cinquantesimo compleanno della consorte Livia Drusilla.

Il nostro studio sul cielo antico ha però evidenziato un secondo elemento che, con ogni probabilità, orientò il principe verso tale scelta: ricostruendo, infatti, il quadro astronomico del 30 gennaio del 9 a.C. alla latitudine di Roma, si dimostra che in tale data il sole sorgeva in congiunzione con la costellazione dell'Acquario mentre il segno del Capricorno appariva poco prima dell'alba (secondo il fenomeno della levata eliaca per cui esso appare trascinare/annunciare il carro del Sole all'arrivo del nuovo giorno per poi fondersi con la luce dell'aurora). Una situazione molto particolare e benaugurale.

Tale fortunata coincidenza ci induce a ritenere che la scelta del giorno in cui celebrare la festa della Pace fu scelta dopo un accorto studio astronomico, finalizzato a garantire la fortuna dell'evento e la presenza di auspici favorevoli.

Il quadro astronomico/astrologico fin qui tracciato è iconicamente sintetizzato da una stele erroneamente attribuita da P. Zanker alle collezioni del Museo Archeologico di Torino (in verità, essa è conservata presso il santuario del Todocco a Pezzolo in valle Uzzone, fig. 25)[112]. Sulla faccia principale sono raffigurati la lupa con i due gemelli e, al di sotto del timpano dell'edicola, due Capricorni affrontati araldicamente ai lati di un disco solare che appare come adagiato su un altare. E' interessante notare che prima del solstizio invernale la costellazione del Capricorno, seppure non visibile, seguiva il sorgere del Sole nella fascia zodiacale e dopo tale data, come suddetto, lo anticipava rendendosi visibile appena prima dell'alba. Dunque l'iconografia della stele sembra voler proprio indicare il "momento" del solstizio invernale e l'importanza del segno del Capricono nell'accompagnare la rinascita del Sole.

La stele di *L. Marius* è stata inquadrata nel filone epigrafico provinciale di estrazione militare e legionaria, dove i simboli cari al principe erano frequentemente riproposti in funzione lealistica e come palese attestazione dell'acquisizione della *romanitas* da parte del defunto[113].

---

[111] MOLLO MEZZENA, 1981, pp. 68-76; EAD., 1988, pp. 74-100.
[112] ZANKE, 2000, pp. 84-91.
[113] CADARIO, 2001, pp. 159-160. La rappresentazione di due Capricorni ai lati di un globo è presente anche su un capitello di anta proveniente dal cosiddetto Tempio di Augusto a Pozzuoli (*Ivi.* pp. 154-155).

Il cippo di L. Mario può essere letto come una vera e propria sintesi iconica del tema astrologico di *Augusta Taurinorum*. Nella scena troviamo una rappresentazione semplificata dello spettacolo naturale che si ripresentava ogni anno in occasione del genetliaco della città che, con ogni probabilità, veniva celebrato presso gli altari cittadini (prima di tutto presso quello del tempio del foro, forse dedicato a Roma e Augusto) quando il sole iniziava a rischiare la città sorgendo alle spalle della collina.

# BIBLIOGRAFIA

TAVOLE ILLUSTRATE

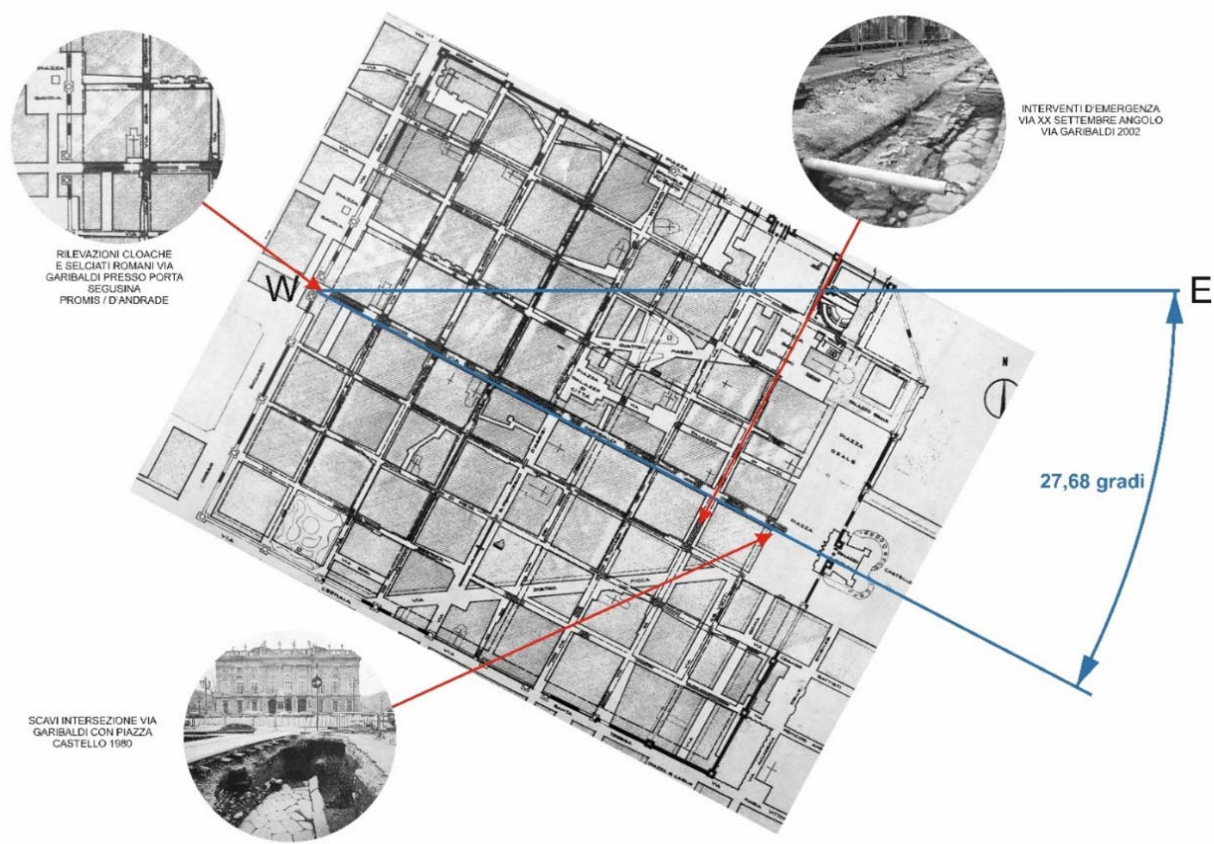

Fig. 2 – Orientamento di *Augusta Taurinorum* con indicazione esemplificativa di parte della documentazione grafica e fotografica a nostra disposizione per ricostruire il tracciato stradale antico.

| SEQUENZA STRATIGRAFICA MURA LATO NORD | | | SEQUENZA STRATIGRAFICA MURA LATO-EST | | |
|---|---|---|---|---|---|
| Teatro con *porticus post scenam* semplice | Età augustea | | edificazione isolato urbano colmatura fossato livellamento suolo | 25-50 d.C. | Livelli prima frequentazione |
| Costruzione cortina muraria | 15-30/40 d.C. | | costruzione cloaca e cortina muraria | 50 – 75 d.C. | Costruzione cortina muraria |
| Adeguamento teatro e costruzione *porticus post scenam* quadrilatera | 30/40-50 d.C. | | primo livello d'uso del sedime stradale | 75-80 d.C. | Formazione della discarica *intra* ed *extra muros* |

Figg 3-4. Cronologia delle mura romane di Torino.

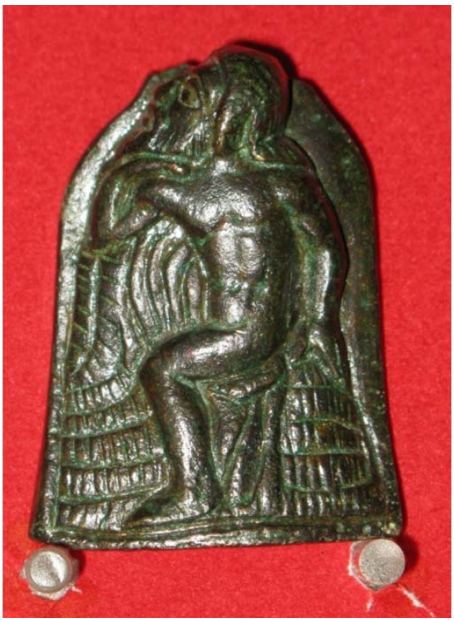

Fig. 5 – Placchetta decorativa raffigurante un *augure* nell'atto di prendere gli *auspicia* conservato nel Museo archeologico di Sarteano.

```
ATILIA ⚬ MVL·L·ONESIME
SIBI·ET·L·AEBVTIO·OPT
ATO·AVG·DEC·AVGVR
CONIVGI · OPTIMO
V·F
```

Fig. 6 – Nell'*Augusta Taurinorum di età imperiale* esercitavano il proprio ruolo degli *àuguri*. Di uno di essi conosciamo il nome grazie ad una epigrafe che era stata trasportata presso il giardino del castello di Moncalieri e nota già a partire dal Cinquecento. Qui, un *L. Aebutius Optatus,* sposato con una liberta, viene ricordato per aver esercitato le funzioni di àugure e di augustale (*CIL* V, 7017).

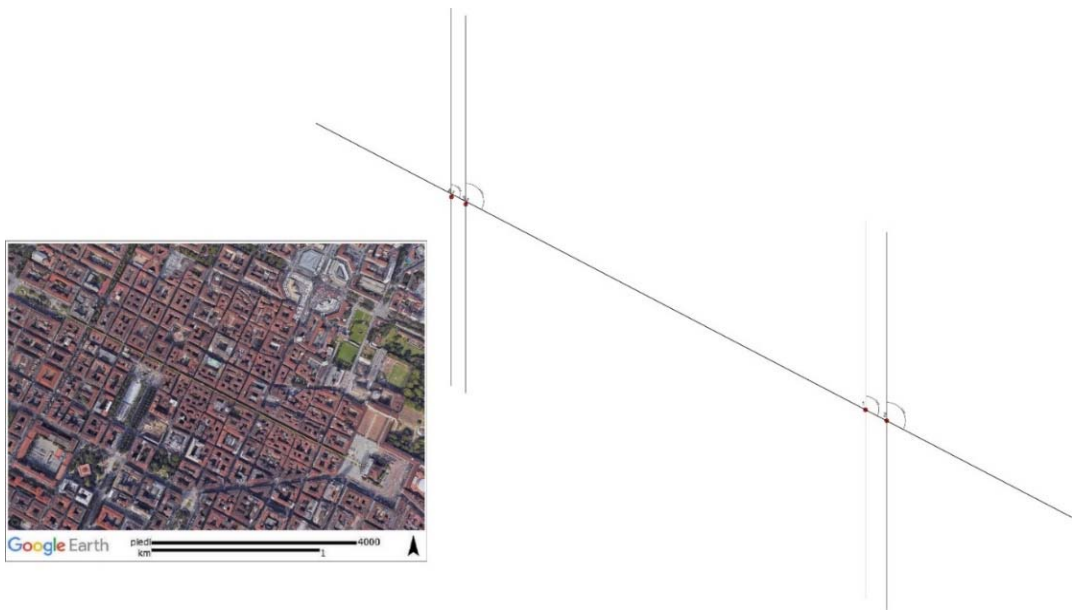

Fig. 7 – Punti di rilevamento dell'azimuth del decumano eseguito con sistema GPS.

| | | | | | | |
|---|---|---|---|---|---|---|
| 1. | p.zza Castello - estr.E | 4991718.6100 | 396461.3091 | 237.2735 | +45° 04' 16.87932000" | +7° 41' 04.84266600" |
| 2. | p.zza Castello - estr.E2 | 4991691.7692 | 396512.4766 | 237.8816 | +45° 04' 16.03667400" | +7° 41' 07.20204600" |
| 3. | p.zza Statuto - ext.W | 4992214.2836 | 395502.6473 | 243.6937 | +45° 04' 32.43083400" | +7° 40' 20.63986200" |
| 4. | p.zza Statuto - ext.W1 | 4992231.8708 | 395468.2464 | 244.0271 | +45° 04' 32.98233000" | +7° 40' 19.05370200" |

Fig. 8 – Rilevazioni GPS dell'orientamento del decumano ottenuti con GPS a multifrequenza.

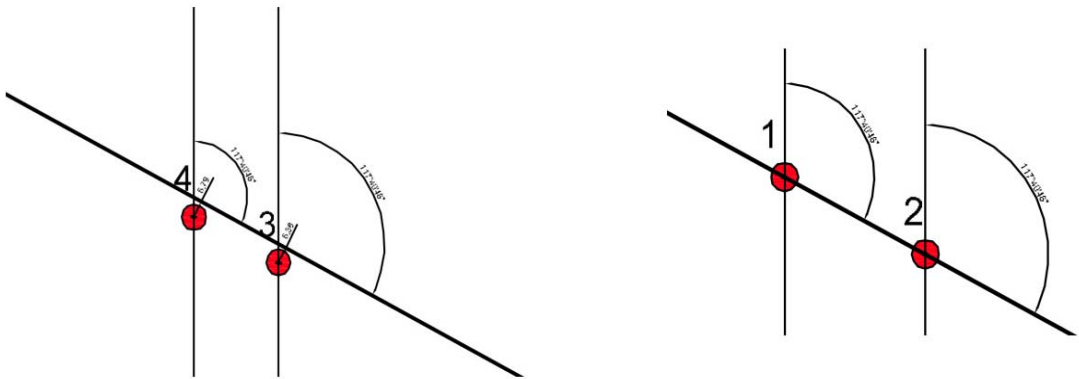

Fig. 9 – L'angolo di decumano misurato rispetto alla direzione N-S in Piazza Statuto.
Fig. 10 – L'angolo di decumano misurato rispetto alla direzione M-S in Piazza Castello.

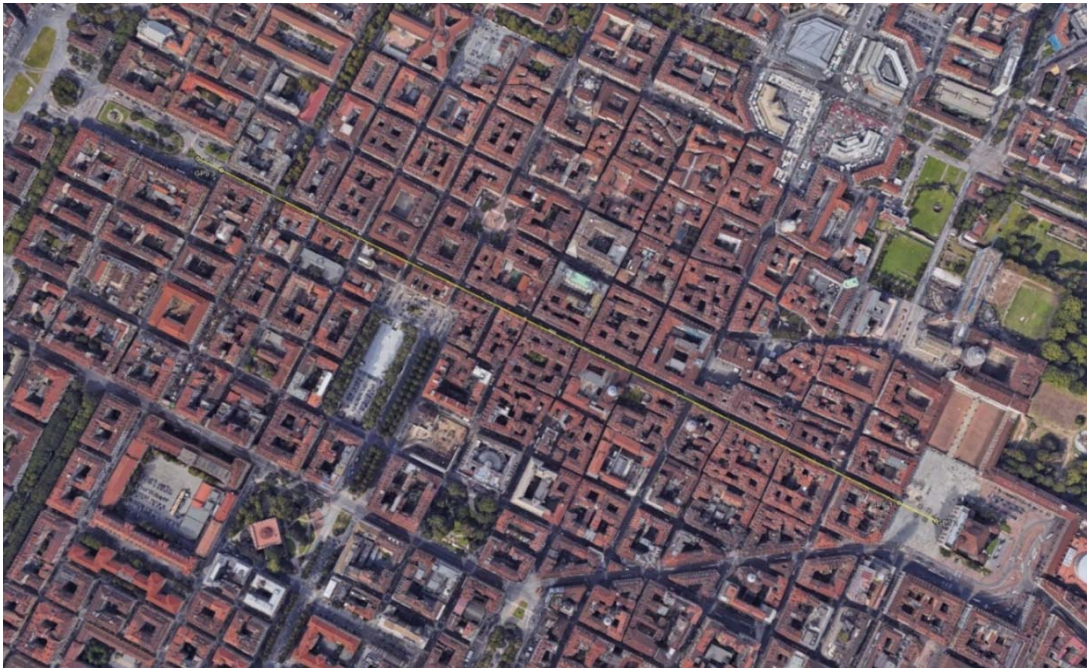

Fig. 11 – L'asse di via Garibaldi/decumano prolungato sino a Piazza Statuto per ridurre l'errore angolare.

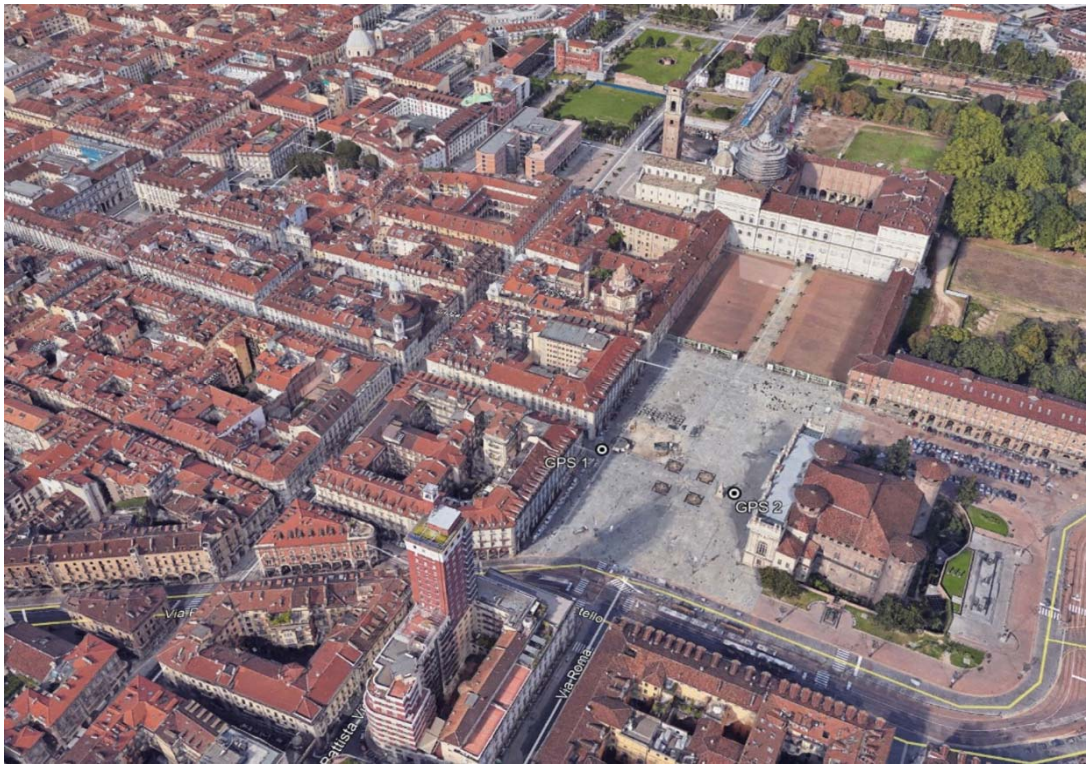

Fig. 12 – Il decumano romano in corrispondenza dall'intersezione tra via Garibaldi e piazza Castello.

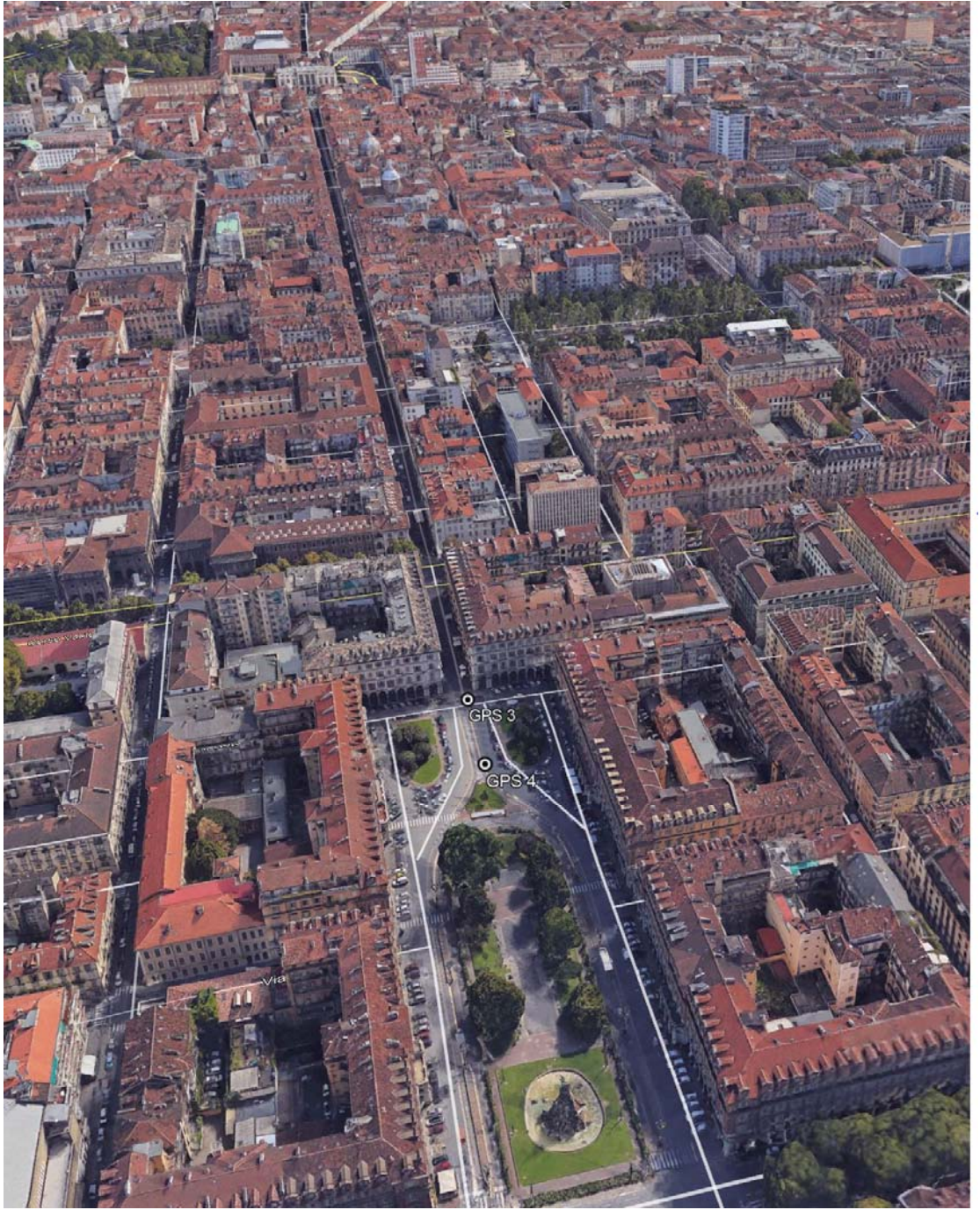

Fig. 13 – L'asse di via Garibaldi/*decumano* e i punti di rilevazione gps ai termini.

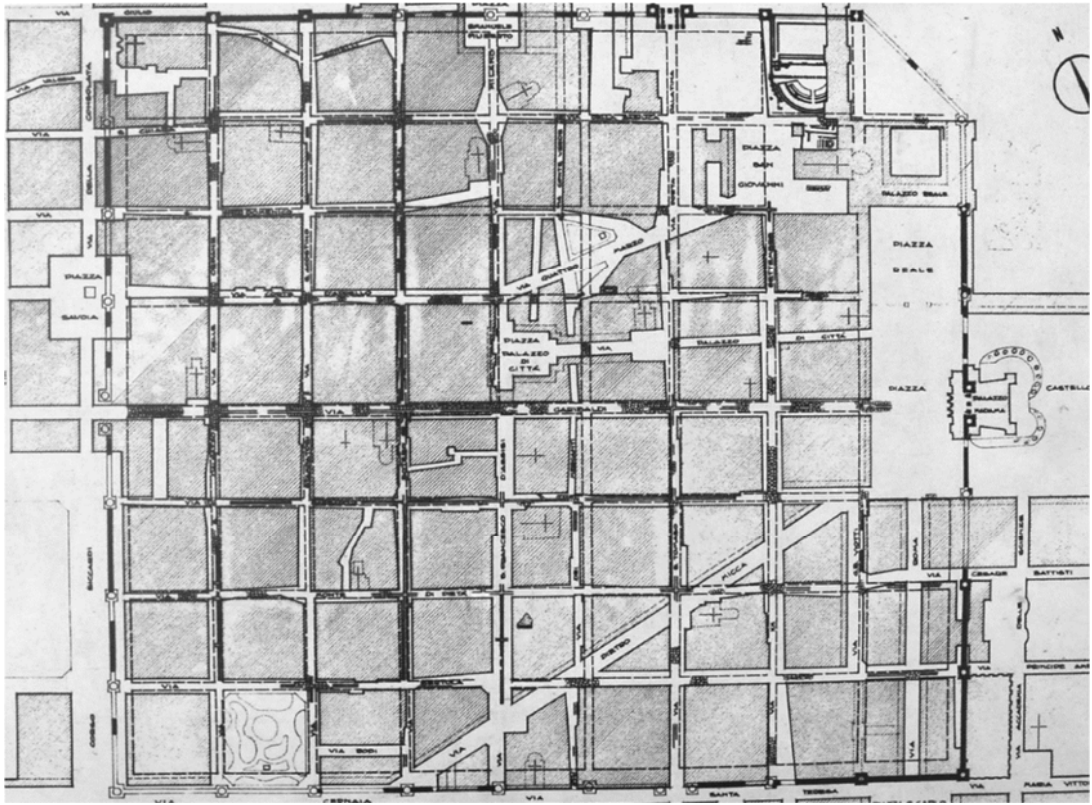

Fig. 14 – La mappa pubblicata da A. D'Andrade con indicazioni dei basolati romani e dei tratti di cloaca romana venuti alla luce nel centro storico di Torino.

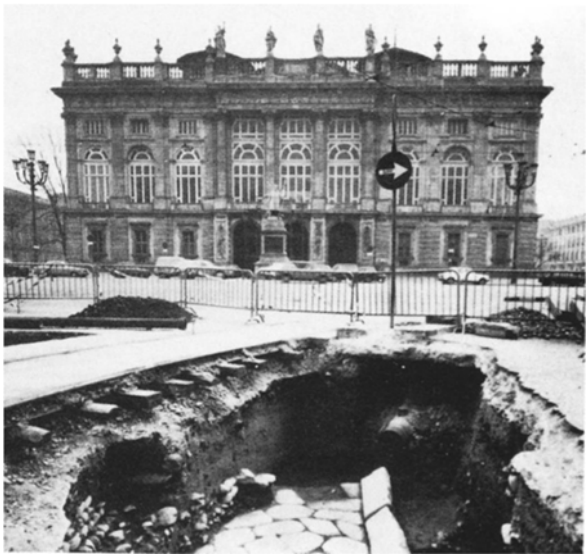

Fig. 15 – Il decumano romano in corrispondenza dell'intersezione tra via Garibaldi e piazza Castello.

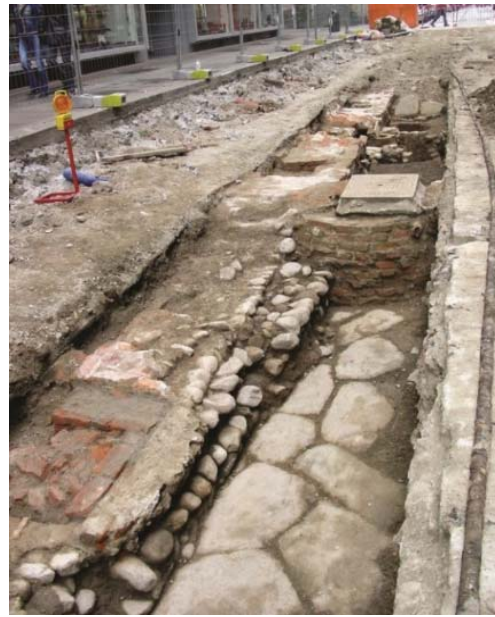

Fig. 16 – Tratto del basolato del cardo romano sotto via XX settembre duranti gli scavi del 2006.

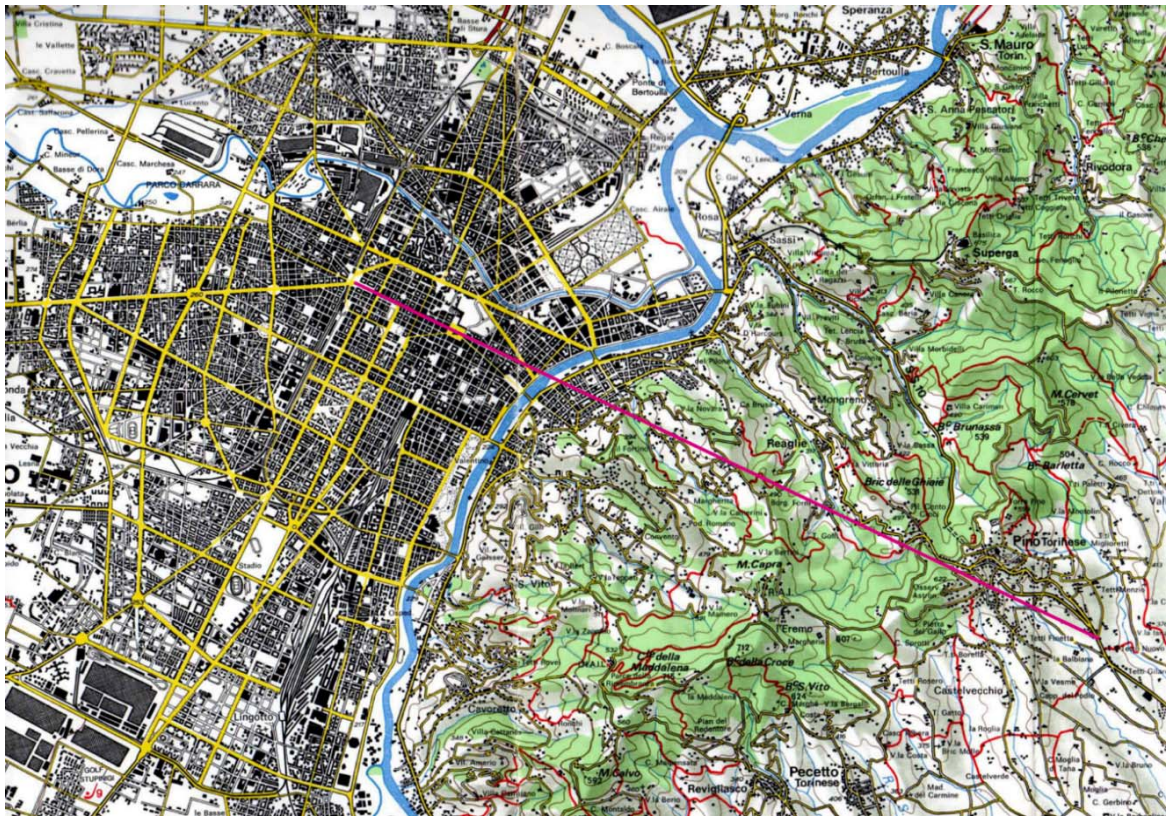

| anno | altezza (gradi) | $A_{\odot crif}$ (gradi) |
|---|---|---|
| -13 | h=1,08 | -27.66 |
|  | h=1,09 | -27.67 |
|  | h=1,10 | -27.68 |
|  | h=1,11 | -27.69 |
|  | h=1,12 | -27.70 |
| -10 | h=1,17 | -27.66 |
|  | h=1,18 | -27.67 |
| -9 | h=1,08 | -27.65 |
|  | h=1,09 | -27.66 |
|  | h=1,10 | -27.67 |
|  | h=1,11 | -27.68 |
|  | h=1,12 | -27.69 |
|  | h=1,13 | -27.70 |
| -6 | h=1,18 | -27.65 |
| -5 | h=1,10 | -27.65 |
|  | h=1,11 | -27.66 |
|  | h=1,12 | -27.68 |
|  | h=1,13 | -27.69 |
|  | h=1,14 | -27.70 |
|  | h=1,15 | -27.71 |
| -1 | h=1,11 | -27.65 |
|  | h=1,12 | -27.66 |
|  | h=1,13 | -27.68 |
|  | h=1,14 | -27.69 |
|  | h=1,15 | -27.70 |
|  | h=1,16 | -27.71 |
| 4 | h=1,13 | -27.66 |
|  | h=1,14 | -27.68 |
|  | h=1,15 | -27.69 |
|  | h=1,16 | -27.70 |
|  | h=1,17 | -27.71 |

Fig.17 – Tracciatura della linea ottica della levata sulla cartografia locale (si ringrazia Renato Pannunzio).

Fig. 18 – Anni ipotizzabili per la fondazione di Torino sulla base del $A_{\odot crif}$ (ampiezza ortiva computata con elevazione e rifrazione) calcolati dal programma attorno ai valori del decumano -27,68° nel 30 gennaio estratti dai valori riportati in appendice B.

Fig. 20. Il fallo, l'aratro e il segno del Capricorno scolpiti sulla Torre dei Balivi di Aosta.

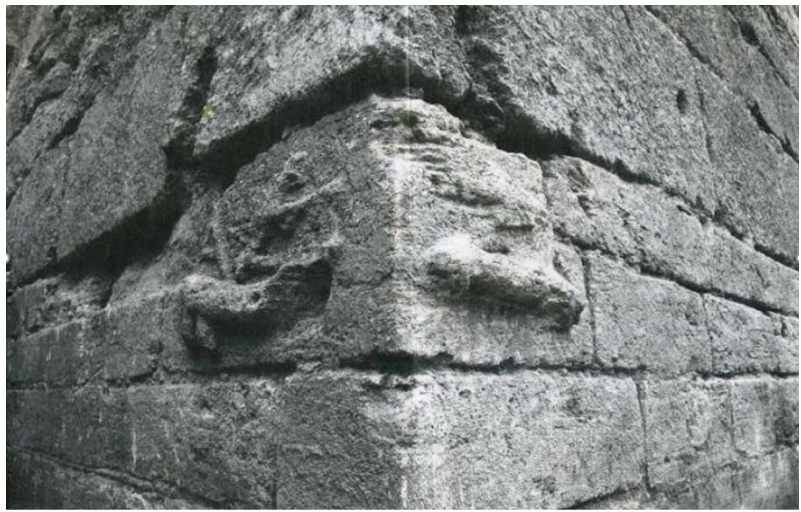

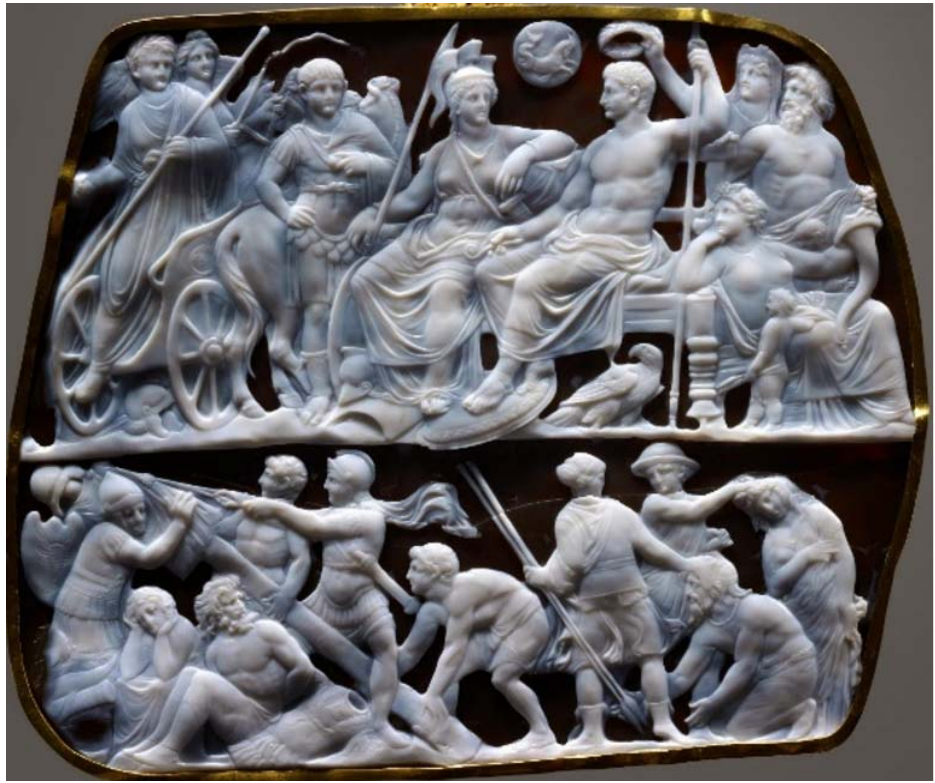

Fig. 21– La Gemma Augustea conservata al Kunsthistorisches Museum di Vienna; in alto, a fianco di Augusto incoronato della personificazione dell'*Oikoumene*, si riconosce un disco che racchiude il segno del Capricorno.

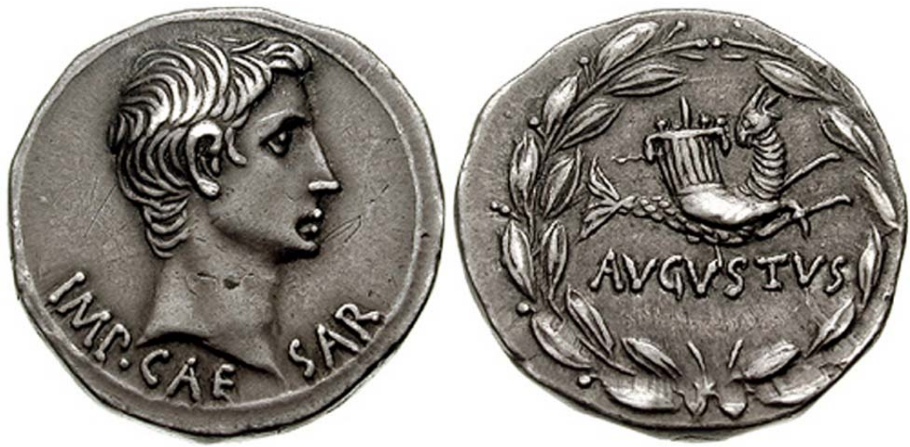

Fig. 22 – Denario di Augusto con rappresentazione del segno zodiacale del Capricorno sul *recto* coniato il 27 a.C.

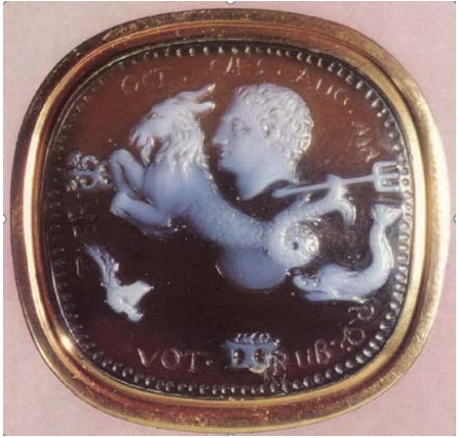

Fig. 23 – Cammeo in onice dell'*Art Institute of Chicago* opera di Dioscuride.

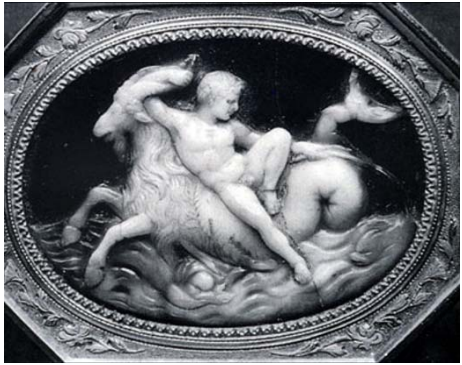

Fig. 24 – Gemma in sardonica conservata nella collezione dell'Hermitage.

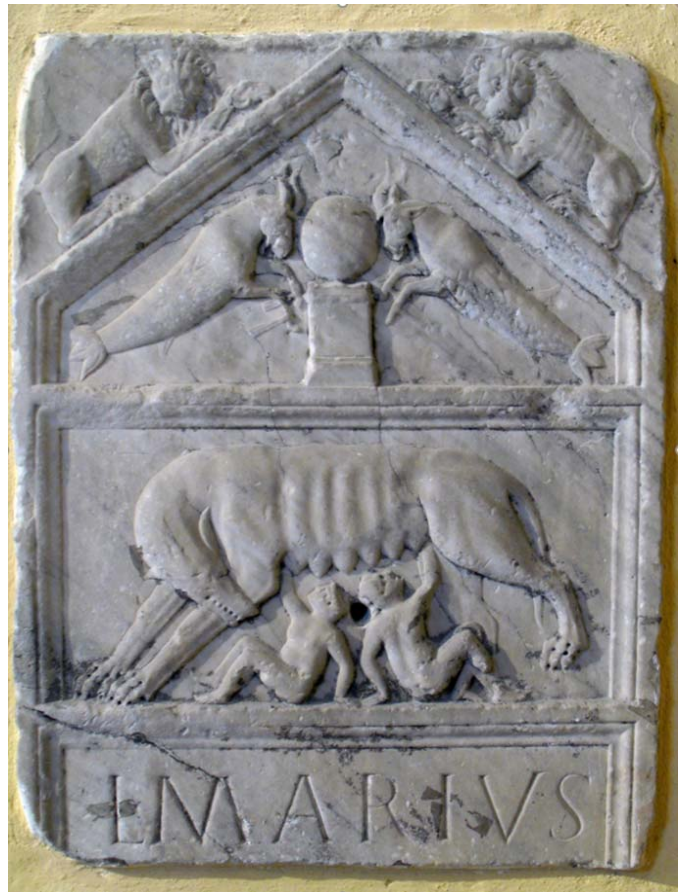

Fig. 25 – Il cippo di L. Mario da Pezzolo con la raffigurazione della Lupa con i Gemelli ed il segno del Capricorno.

**Appendice A: Elementi di base e precisazioni sul calcolo astronomico**

Come è noto nell'astronomia la posizione degli astri in cielo viene determinata mediante regole di trigonometria sferica[114]. Infatti, data la loro enorme distanza, gli oggetti celesti appaiono come proiettati su una sfera e le loro posizioni sono individuate attraverso misure di coordinate angolari.
Per un osservatore terrestre il moto locale del Sole[115] può rappresentarsi opportunamente per mezzo del primo sistema di coordinate altazimutali[116], ovvero dell'Azimut (A)[117] e dell'Altezza (h)[118], utilizzando come riferimenti l'Orizzonte Astronomico[119] e il Meridiano Celeste (quest'ultimo passante per lo Zenit (Z) ed orientato secondo la direzione Nord-Sud). In generale, l'Orizzonte Astronomico non coincide con l'orizzonte sensibile e, tuttavia, può essere facilmente determinato con un filo a piombo essendo perpendicolare alla direzione dell'accelerazione di gravità.

Diversi fattori concorrono a disegnare l'analemma del Sole ad una determinata latitudine terrestre. Primo tra questi è il fatto che il moto apparente del Sole lungo l'eclittica non avviene in modo costante per effetto della seconda legge di Keplero, secondo la quale la Terra ruota più velocemente al perielio rispetto all'afelio. Segue la variazione dell'obliquità dell'eclittica, calcolata in prima approssimazione secondo la formula di Newcomb[120].

Per tale ragione, in Astronomia si usa definire come 'Sole medio' un punto immaginato al suo centro, per mezzo del quale è possibile calcolare in via teorica la posizione dell'astro sulla sfera celeste. In un anno (pari a 365,25 giorni solari medi) il Sole percorre sull'eclittica 360°, quindi poco meno di un grado al giorno. Il Sole segue, al contempo, un moto a ritroso lungo l'eclittica e culmina in ritardo nel passaggio al Meridiano Celeste rispetto ad orologi

---

[114] In base alle necessità di studio, i sistemi di coordinate si distinguono in funzione della scelta dell'origine (il meridiano di riferimento, il punto gamma, il centro del Sole, il centro Galattico, etc..) e del piano di riferimento (l'orizzonte locale astronomico, il piano equatoriale terrestre, quello dell'eclittica o galattico, etc..).
[115] In Astronomia si usa indicare il Sole con il simbolo ☉.
[116] Dato che la rotazione terrestre giornaliera altera le coordinate altazimutali degli astri in continuazione, essi apparentemente descrivono, da Est verso Ovest, sulla volta Celeste, degli archi di cerchio non paralleli all'Orizzonte (salvo che al Polo Nord o Sud) e la massima altezza che raggiungono sull'orizzonte è denominata Culminazione Massima o Superiore. In particolare, il moto del Sole rispetto alle stelle fisse appare avvenire lungo un cerchio massimo detto Eclittica. Inoltre, gli astri hanno posizioni e tempi di passaggio differenti a seconda del luogo di osservazione.
[117] Dall'arabo As-Samt che significa direzione, rappresenta l'arco di Orizzonte compreso fra il punto cardinale Nord e la proiezione dell'oggetto osservato P sull'orizzonte. In questo caso è detto geodetico, mentre in Astronomia è diffusa la misura degli Azimut a partire dal punto cardinale Sud, positivamente verso Ovest e negativamente verso Est.
[118] Essa viene misurata positivamente in gradi dal punto P sull'Orizzonte lungo il Cerchio Verticale fino ad incontrare l'Astro osservato oppure dallo Zenit (Z) (quindi per "Zenit" si intende anche l'angolo complementare dell'altezza).
[119] Identificato dall'ideale cerchio massimo risultante dall'intersezione della sfera celeste con il piano passante per l'osservatore.
[120] LATTANZI, PANNUNZIO, 2006, pp. 46-48

di tempo siderali[121].

Per ottenere un giorno di durata costante associato al moto del Sole, gli astronomi hanno adottato due punti fittizi, quello del 'Sole eclitticale medio' e quello del 'Sole equatoriale medio'. Il primo si sposta uniformemente sull'eclittica con la velocità media del Sole e coincide con 'il vero' al perielio (a gennaio) ed all'afelio (intorno al 4 luglio). Il secondo si sposta, invece, uniformemente sull'equatore celeste con la velocità costante del Sole eclitticale medio e coincide con esso ai due equinozi.

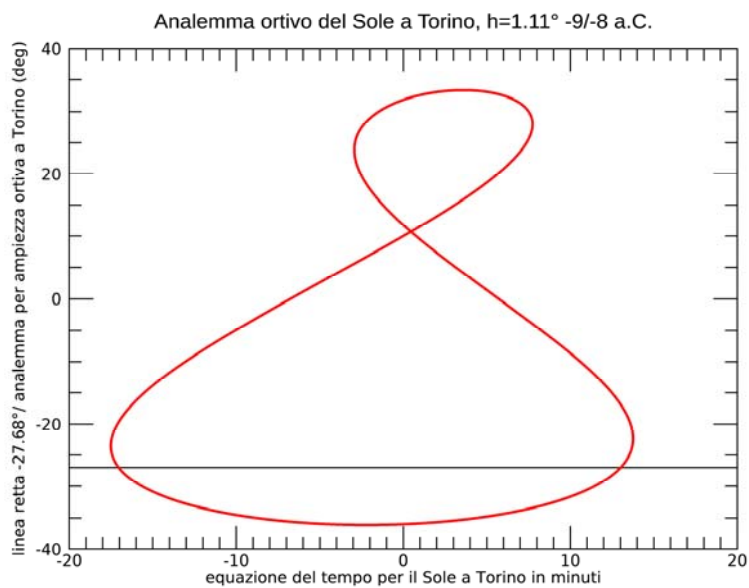

Un altro modo per stimare la posizione di un astro è considerare, anziché l'Azimut, il cosiddetto angolo orario H, ovvero l'arco di equatore contato a partire dal punto di intersezione tra Meridiano Celeste e l'equatore celeste, detto mezzocielo, fino alla proiezione dell'astro sull'equatore celeste considerato ora come piano di riferimento.
H si conta in ora, minuti e secondi in senso orario, in direzione ovest, in modo che cresca con la rotazione terrestre, ovvero con il tempo ad intervalli regolari di 24h. Praticamente conta quanto tempo è passato dal passaggio al meridiano dell'astro. La seconda coordinata in questo caso è la declinazione, contata a partire dall'equatore celeste lungo il cerchio massimo passante per l'astro e la sua proiezione su di esso.

---

[121] Esistono diversi modi di misurare il tempo locale. In astronomia si usa il giorno siderale definito come l'intervallo di tempo in cui, per effetto della rotazione apparente della sfera celeste, una stella compie un giro completo attorno al polo. L'ora siderale zero è momento in cui l'astro culmina. Di norma non ci si riferisce ad una stella in particolare, ma al punto vernale, indicante l'intersezione del piano dell'eclittica con l'equatore celeste, che, come tutte le stelle, partecipa al moto di rotazione della sfera celeste. Siccome esso non rappresenta un oggetto fisico, il tempo siderale è definito facendo riferimento in sostanza all'angolo orario del Sole all'equinozio di primavera, essendo infatti la posizione in cui il Sole attraversa l'equatore celeste nel suo moto verso l'afelio. Si noti che le posizioni reciproche del Sole e del punto variano durante il corso dell'anno, per cui il Sole culmina in istanti diversi di tempo siderale nei vari giorni dell'anno. Anticamente l'equinozio di primavera segnava in genere l'inizio di un nuovo anno.

Pertanto se trattiamo il Sole alla pari di un astro abbiamo due sistemi di coordinate per tracciare il suo moto locale in cielo. Per il nostro studio, in particolare, è importante individuare in quali date l'angolo di levata del Sole corrisponda al valore dell'angolo del decumano misurato, ovvero l'azimut (o ampiezza ortiva rispetto ad est) del Sole nascente al momento della fondazione di Torino; il corrispondente angolo orario dovrebbe indicare quanto tempo manchi al passaggio al meridiano (ore 12).

Alla latitudine di Torino (45,07°) la formula da utilizzare per l'angolo orario ad una determinate altezza è la seguente:

$$H = \frac{180}{\pi} \arccos\left( \frac{\sin(-h)}{\cos \lambda_{To} \cos \delta} - \tan \lambda_{To} \tan \delta \right) \quad (3)$$

Sottolineamo ancora che la declinazione del Sole adottata tiene conto del suo moto "vero".

Per quanto concerne la fondazione di *Augusta Taurinorum*, l'angolo orario risulta pari a $7^h, 41$, ovvero 7 h 24 min 36 s (tavola A1) ed indica l'orario del sorgere del bordo del Sole Vero per h =1.11° (il risultato non cambia significativamente per i restanti valori della fig. 18).

Per il calcolo dell'angolo orario medio locale $H_m$ occorre tener conto della correzione per l'equazione del tempo pari a -16,82 minuti (ovvero ET=$H_v$ - $H_m$, calcolata dal programma IDL per il 30 gennaio del 9 a.C.), per cui risulterebbe $H_m$= $7^h$,69 = 7 h 41 min 24 s. A questo orario bisognerebbe aggiungere qualche minuto se si volesse considerare il levare del centro del Sole e non del suo bordo. Rispetto al 30 Gennaio 2019 l'equazione del tempo indica uno scarto di circa 3min 34 s (ET= -13.25 minuti, si vedano i grafici sotto riportati per l'analemma ortivo).

Se, invece, si considerasse la correzione per la longitudine di Torino pari a 29,34 minuti, l'ora della levata rispetto al tempo del fuso vero sarebbe 7 h 53 min 56 s, valore confrontabile con quelli odierni. Rispetto ad un Sole medio bisognerebbe aggiungere ancora 16,82 minuti, se si volesse esprimere l'ora del levare nelle unità di tempo civile che usiamo oggigiorno.

**Tavola A1: angoli orari per h=1.11°**

| JD | data civile | declinazione | ampiezza ortiva (A) | angolo orario |
|---|---|---|---|---|
| 1716704.000000 | 01/30/ -13 | -18.327623° | -27.694077° | 7$^h$.410607 |
| 1716991.000000 | 11/12/ -13 | -18.297289° | -27.648026° | 7$^h$.408175 |
| **1718165.000000** | **01/30/ -9** | **-18.320064°** | **-27.682601°** | **7$^h$.410001** |
| 1718452.000000 | 11/12/ -9 | -18.304491° | -27.658959° | 7$^h$.408752 |
| 1719626.000000 | 01/30/ -5 | -18.310559° | -27.668170° | 7$^h$.409239 |
| 1719913.000000 | 11/12/ -5 | -18.313476° | -27.672599° | 7$^h$.409473 |
| 1721087.000000 | 01/30/ -1 | -18.313476 ° | -27.653578° | 7$^h$.408468 |
| 1721374.000000 | 11/12/ -1 | -18.322035° | -27.685592° | 7$^h$.410159 |

**Grafici delle ampiezze ortive in funzione dell'equazione del tempo**

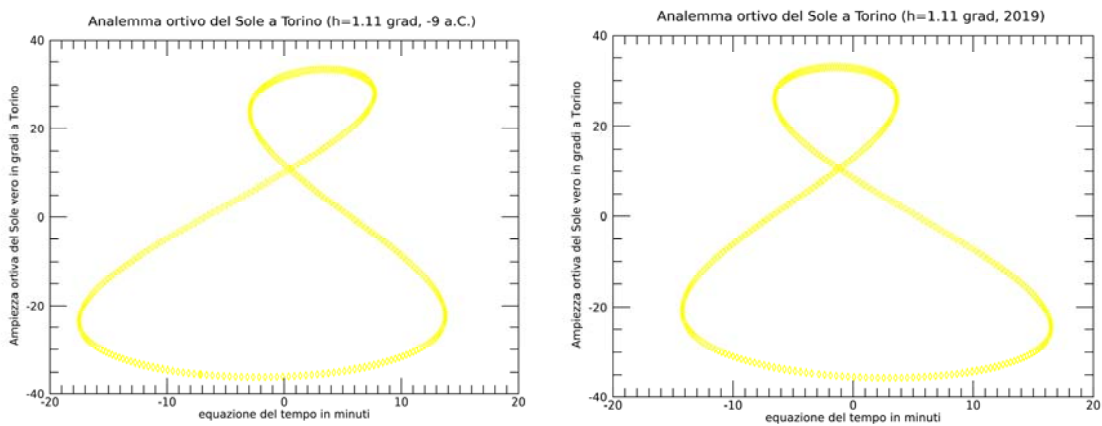

Precisiamo che, qualora si volesse calcolare l'ampiezza ortiva del Sole utilizzando l'angolo orario, si trovano valori confrontabili con quelli del decumano (-27,68°) verso il 10 e il 30 dicembre, ma solo nel caso si tenga conto della rifrazione e della collina; tuttavia, essi non possono essere significativi in quanto l'azimut del Sole in quel periodo passa da 124° a 122° circa, corrispondenti ad un'ampiezza ortiva tra -34° a -32°, ovvero con uno scarto minimo di almeno 4° rispetto a quello misurato in via Garibaldi (a maggior ragione per il solstizio invernale: circa 126° di azimut, quindi un'ampiezza ortiva di circa -36°).

Di seguito vengono riportati i grafici di confronto con il valore dell'azimut del decumano di circa 26°. Notare che nel caso di un azimut del decumano pari a -27,68° per h=0 non vi è nessun valore coincidente con il solstizio invernale per il Sole Vero, risultato che evidenzia l'erronea deduzione allorché si utilizzi un moto apparente del Sole semplificato.

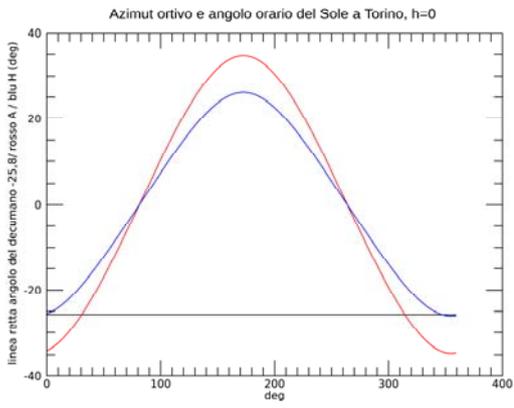
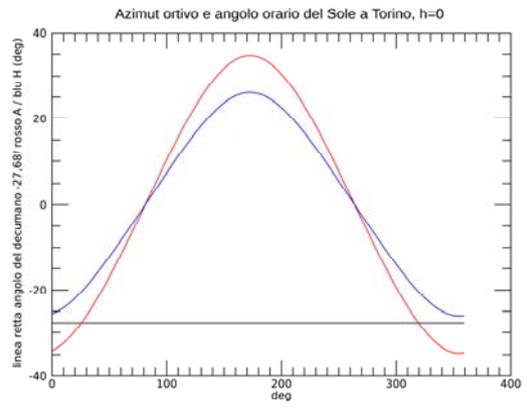
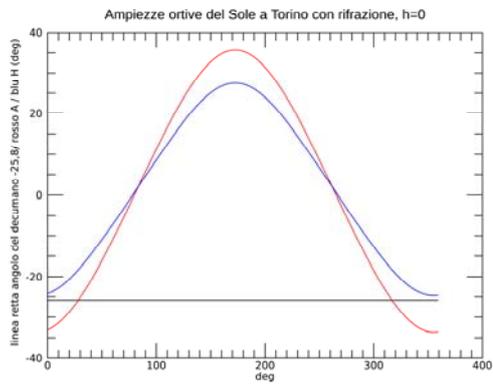
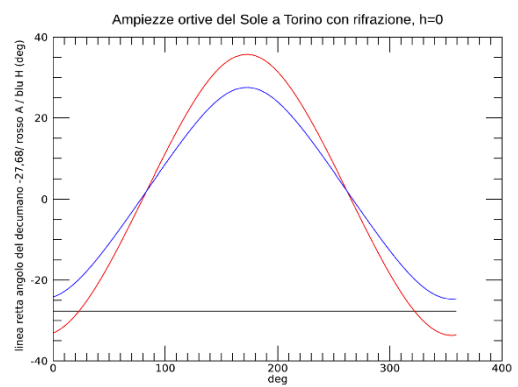
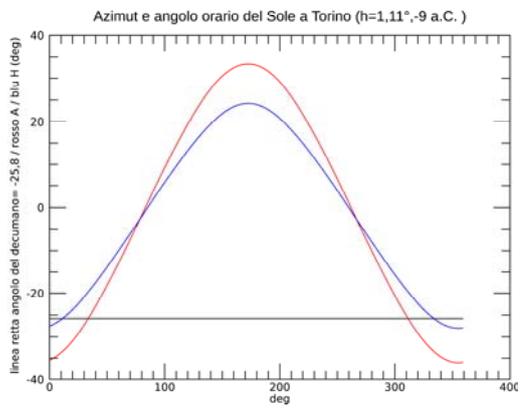
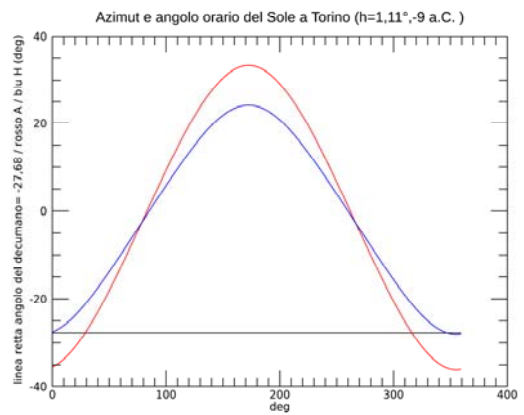

**Tavola A2: risultati relativi al mese di dicembre per l'angolo orario H**

| JD | data civile | declinazione | 90°-H | angolo orario |
|---|---|---|---|---|
| 1717022.000000 | 12/13/ -13 | -23.448285° | -27.690764° | $7^h.846051$ |
| 1717038.000000 | 12/29/ -13 | -23.468407° | -27.718132° | $7^h.847875$ |
| 1717387.000000 | 12/13/ -12 | -23.433205° | -27.670264° | $7^h.844684$ |
| 1717403.000000 | 12/29/ -12 | -23.482197° | -27.736897° | $7^h.849126$ |
| 1717753.000000 | 12/14/ -11 | -23.478171° | -27.731417° | $7^h.848761$ |
| 1718118.000000 | 12/14/ -10 | -23.464067° | -27.712228° | $7^h.847482$ |
| 1718134.000000 | 12/30/ -10 | -23.452029° | -27.695854° | $7^h.846390$ |
| 1718483.000000 | 12/13/ -9 | -23.449565° | -27.692504° | $7^h.846167$ |
| 1718499.000000 | 12/29/ -9 | -23.466212° | -27.715146° | $7^h.847676$ |
| 1718848.000000 | 12/13/ -8 | -23.434664° | -27.672247° | $7^h.844816$ |
| 1718864.000000 | 12/29/ -8 | -23.479939° | -27.733825° | $7^h.848922$ |
| 1719214.000000 | 12/14/ -7 | -23.479565° | -27.733315° | $7^h.848888$ |
| 1719230.000000 | 12/30/ -7 | -23.434831° | -27.672473° | $7^h.844832$ |
| 1719579.000000 | 12/14/ -6 | -23.465617° | -27.714336° | $7^h.847622$ |
| 1719595.000000 | 12/30/ -6 | -23.449438° | -27.692332° | $7^h.846155$ |
| 1719944.000000 | 12/13/ -5 | -23.451246° | -27.694790° | $7^h.846319$ |
| 1719960.000000 | 12/29/ -5 | -23.463618° | -27.711617° | $7^h.847441$ |

**Tavola A3: risultati relativi al solstizio invernale**

| JD | data civile | declinazione | ampiezza ortiva (A) | angolo orario |
|---|---|---|---|---|
| 1717030.000000 | 12/21/ -13 | -23.696130 ° | -36.056161 ° | 61.971052° |
| 1717032.000000 | 12/23/ -13 | -23.683754° | -36.036310° | 61.987996° |
| 1717395.000000 | 12/21/ -12 | -23.695480 ° | -36.055118 ° | 61.971943° |
| 1717397.000000 | 12/23/ -12 | -23.686729 ° | -36.041082 ° | 61.983924° |
| 1717760.000000 | 12/21/ -11 | -23.694390 ° | -36.053369 ° | 61.973435° |
| 1717762.000000 | 12/23/ -11 | -23.689258 ° | -36.045139 ° | 61.980461° |
| 1718125.000000 | 12/21/ -10 | -23.692864 ° | -36.050922 ° | 61.975525° |
| 1718127.000000 | 12/23/ -10 | -23.691344° | -36.048484° | 61.977606° |
| 1718491.000000 | 12/21/ -9 | -23.695671° | -36.055424° | 61.971681° |
| 1718495.000000 | 12/23/ -9 | -23.640279° | -35.966603° | 62.047468° |

**Appendice B: risultati numerici del calcolo astronomico**

**Tabella α:** ampiezze ortive $A_{\odot crif}$ (ultima colonna) attorno al valore del decumano -27,68°± 0,07° dal 27 a.C. al 4 d.C. calcolate tenendo conto della rifrazione e a partire dall'altezza $h = 1.06°$ incrementata di 0,01° fino a $h = 1.18°$ (prima colonna). Nella terza colonna sono riportate le conversioni dell'anno Giuliano (seconda colonna) nelle corrispondenti dati civili. Nella quarta colonna è riportato il valore della declinazione del Sole Vero.

| altezza | data giuliana | giorno, mese, anno | $\delta_\odot$ (gradi) | $A_{\odot crif}$ (gradi) |
|---|---|---|---|---|
| h=1,06° | 1711878.000000 | 11/13/ -27 | -18.400427 | -27.747500 |
| | 1712243.000000 | 11/13/ -26 | -18.337816 | -27.652457 |
| | 1712321.000000 | 01/30/ -25 | -18.352704 | -27.675052 |
| | 1713704.000000 | 11/13/ -22 | -18.346936 | -27.666298 |
| | 1713782.000000 | 01/30/ -21 | -18.342843 | -27.660085 |
| | 1715165.000000 | 11/13/ -18 | -18.354696 | -27.678076 |
| | 1715243.000000 | 01/30/ -17 | -18.334429 | -27.647316 |
| | 1715608.000000 | 01/29/ -16 | -18.397400 | -27.742904 |
| | 1716626.000000 | 11/13/ -14 | -18.360765 | -27.687287 |
| | 1716704.000000 | 01/30/ -13 | -18.327623 | -27.636988 |
| | 1717069.000000 | 01/29/ -12 | -18.390686 | -27.732710 |
| | 1718087.000000 | 11/13/ -10 | -18.367444 | -27.697426 |
| | 1718165.000000 | 01/30/ -9 | -18.320064 | -27.625518 |
| | 1718530.000000 | 01/29/ -8 | -18.382664 | -27.720531 |
| | 1719548.000000 | 11/13/ -6 | -18.376072 | -27.710524 |
| | 1719626.000000 | 01/30/ -5 | -18.310559 | -27.611095 |
| | 1719913.000000 | 11/12/ -5 | -18.313476 | -27.615521 |
| | 1719991.000000 | 01/29/ -4 | -18.372923 | -27.705742 |
| | 1721009.000000 | 11/13/ -2 | -18.384934 | -27.723977 |
| | 1721374.000000 | 11/12/ -1 | -18.322035 | -27.628508 |
| | 1721452.000000 | 01/29/ 1 | -18.363730 | -27.691788 |
| | 1722470.000000 | 11/13/ 3 | -18.391944 | -27.734620 |
| | 1722835.000000 | 11/12/ 4 | -18.328618 | -27.638499 |
| h=1,07° | 1712243.000000 | 11/13/ -26 | -18.337816 | -27.663572 |
| | 1712321.000000 | 01/30/ -25 | -18.352704 | -27.686470 |
| | 1713704.000000 | 11/13/ -22 | -18.346936 | -27.677715 |
| | 1713782.000000 | 01/30/ -21 | -18.342843 | -27.671501 |
| | 1715165.000000 | 11/13/ -18 | -18.354696 | -27.689493 |
| | 1715243.000000 | 01/30/ -17 | -18.334429 | -27.658731 |
| | 1716626.000000 | 11/13/ -14 | -18.360765 | -27.698706 |
| | 1716704.000000 | 01/30/ -13 | -18.327623 | -27.648402 |
| | 1717069.000000 | 01/29/ -12 | -18.390686 | -27.744133 |
| | 1718087.000000 | 11/13/ -10 | -18.367444 | -27.708846 |
| | 1718165.000000 | 01/30/ -9 | -18.320064 | -27.636931 |
| | 1718452.000000 | 11/12/ -9 | -18.304491 | -27.613299 |
| | 1718530.000000 | 01/29/ -8 | -18.382664 | -27.731953 |
| | 1719548.000000 | 11/13/ -6 | -18.376072 | -27.721945 |
| | 1719626.000000 | 01/30/ -5 | -18.310559 | -27.622505 |
| | 1719913.000000 | 11/12/ -5 | -18.313476 | -27.626933 |
| | 1719991.000000 | 01/29/ -4 | -18.372923 | -27.717163 |
| | 1721009.000000 | 11/13/ -2 | -18.384934 | -27.735400 |
| | 1721374.000000 | 11/12/ -1 | -18.322035 | -27.639920 |
| | 1721452.000000 | 01/29/ 1 | -18.363730 | -27.703207 |
| | 1722470.000000 | 11/13/ 3 | -18.391944 | -27.746044 |
| | 1722835.000000 | 11/12/ 4 | -18.328618 | -27.649912 |
| h=1,08° | 1712243.000000 | 11/13/ -26 | -18.337816 | -27.675290 |
| | 1712321.000000 | 01/30/ -25 | -18.352704 | -27.697890 |
| | 1713704.000000 | 11/13/ -22 | -18.346936 | -27.689133 |
| | 1713782.000000 | 01/30/ -21 | -18.342843 | -27.682920 |
| | 1715165.000000 | 11/13/ -18 | -18.354696 | -27.700913 |
| | 1715243.000000 | 01/30/ -17 | -18.334429 | -27.670148 |
| | 1716626.000000 | 11/13/ -14 | -18.360765 | -27.710127 |
| | 1716704.000000 | 01/30/ -13 | -18.327623 | -27.659817 |
| | 1716991.000000 | 11/12/ -13 | -18.297289 | -27.613782 |
| | 1718087.000000 | 11/13/ -10 | -18.367444 | -27.720268 |
| | 1718165.000000 | 01/30/ -9 | -18.320064 | -27.648345 |
| | 1718452.000000 | 11/12/ -9 | -18.304491 | -27.624710 |
| | 1718530.000000 | 01/29/ -8 | -18.382664 | -27.743378 |
| | 1719548.000000 | 11/13/ -6 | -18.376072 | -27.733369 |
| | 1719626.000000 | 01/30/ -5 | -18.310559 | -27.633918 |
| | 1719913.000000 | 11/12/ -5 | -18.313476 | -27.638346 |
| | 1719991.000000 | 01/29/ -4 | -18.372923 | -27.728586 |
| | 1721009.000000 | 11/13/ -2 | -18.384934 | -27.746825 |
| | 1721087.000000 | 01/30/ -1 | -18.300947 | -27.619332 |
| | 1721374.000000 | 11/12/ -1 | -18.322035 | -27.651335 |
| | 1721452.000000 | 01/29/ 1 | -18.363730 | -27.714628 |
| | 1722835.000000 | 11/12/ 4 | -18.328618 | -27.661328 |
| h=1,09° | 1711956.000000 | 01/30/ -26 | -18.290132 | -27.614333 |
| | 1712243.000000 | 11/13/ -26 | -18.337816 | -27.686709 |
| | 1712321.000000 | 01/30/ -25 | -18.352704 | -27.709312 |
| | 1713704.000000 | 11/13/ -22 | -18.346936 | -27.700554 |
| | 1713782.000000 | 01/30/ -21 | -18.342843 | -27.694340 |
| | 1715165.000000 | 11/13/ -18 | -18.354696 | -27.712336 |
| | 1715243.000000 | 01/30/ -17 | -18.334429 | -27.681566 |
| | 1715530.000000 | 11/12/ -17 | -18.291286 | -27.616085 |
| | 1716626.000000 | 11/13/ -14 | -18.360765 | -27.721550 |
| | 1716704.000000 | 01/30/ -13 | -18.327623 | -27.671235 |
| | 1716991.000000 | 11/12/ -13 | -18.297289 | -27.625194 |
| | 1718087.000000 | 11/13/ -10 | -18.367444 | -27.731693 |
| | 1718165.000000 | 01/30/ -9 | -18.320064 | -27.659761 |
| | 1718452.000000 | 11/12/ -9 | -18.304491 | -27.636124 |
| | 1719548.000000 | 11/13/ -6 | -18.376072 | -27.744795 |
| | 1719626.000000 | 01/30/ -5 | -18.310559 | -27.645333 |
| | 1719913.000000 | 11/12/ -5 | -18.313476 | -27.649761 |
| | 1719991.000000 | 01/29/ -4 | -18.372923 | -27.740011 |
| | 1721087.000000 | 01/30/ -1 | -18.300947 | -27.630745 |
| | 1721374.000000 | 11/12/ -1 | -18.322035 | -27.662752 |
| | 1721452.000000 | 01/29/ 1 | -18.363730 | -27.726052 |
| | 1722548.000000 | 01/30/ 4 | -18.293230 | -27.619034 |
| | 1722835.000000 | 11/12/ 4 | -18.328618 | -27.672746 |

| h=1,10° | 1711956.000000 | 01/30/ -26 | -18.290132 | -27.625747 |
|---|---|---|---|---|
| | 1712243.000000 | 11/13/ -26 | -18.337816 | -27.698131 |
| | 1712321.000000 | 01/30/ -25 | -18.352704 | -27.720736 |
| | 1713417.000000 | 01/30/ -22 | -18.280233 | -27.610723 |
| | 1713704.000000 | 11/13/ -22 | -18.346936 | -27.711977 |
| | 1713782.000000 | 01/30/ -21 | -18.342843 | -27.705762 |
| | 1714069.000000 | 11/12/ -21 | -18.284016 | -27.616465 |
| | 1715165.000000 | 11/13/ -18 | -18.354696 | -27.723760 |
| | 1715243.000000 | 01/30/ -17 | -18.334429 | -27.692988 |
| | 1715530.000000 | 11/12/ -17 | -18.291286 | -27.627499 |
| | 1716626.000000 | 11/13/ -14 | -18.360765 | -27.732975 |
| | 1716704.000000 | 01/30/ -13 | -18.327623 | -27.682655 |
| | 1716991.000000 | 11/12/ -13 | -18.297289 | -27.636609 |
| | 1718087.000000 | 11/13/ -10 | -18.367444 | -27.743119 |
| | 1718165.000000 | 01/30/ -9 | -18.320064 | -27.671180 |
| | 1718452.000000 | 11/12/ -9 | -18.304491 | -27.647541 |
| | 1719626.000000 | 01/30/ -5 | -18.310559 | -27.656750 |
| | 1719913.000000 | 11/12/ -5 | -18.313476 | -27.661179 |
| | 1721087.000000 | 01/30/ -1 | -18.300947 | -27.642161 |
| | 1721374.000000 | 11/12/ -1 | -18.322035 | -27.674171 |
| | 1721452.000000 | 01/29/ 1 | -18.363730 | -27.737478 |
| | 1722548.000000 | 01/30/ 4 | -18.293230 | -27.630449 |
| | 1722835.000000 | 11/12/ 4 | -18.328618 | -27.684166 |
| h=1,11° | 1711956.000000 | 01/30/ -26 | -18.290132 | -27.637163 |
| | 1712243.000000 | 11/13/ -26 | -18.337816 | -27.709554 |
| | 1712321.000000 | 01/30/ -25 | -18.352704 | -27.732162 |
| | 1712608.000000 | 11/12/ -25 | -18.274982 | -27.614168 |
| | 1713417.000000 | 01/30/ -22 | -18.280233 | -27.622138 |
| | 1713704.000000 | 11/13/ -22 | -18.346936 | -27.723403 |
| | 1713782.000000 | 01/30/ -21 | -18.342843 | -27.717187 |
| | 1714069.000000 | 11/12/ -21 | -18.284016 | -27.627880 |
| | 1715165.000000 | 11/13/ -18 | -18.354696 | -27.735186 |
| | 1715243.000000 | 01/30/ -17 | -18.334429 | -27.704411 |
| | 1715530.000000 | 11/12/ -17 | -18.291286 | -27.638915 |
| | 1716626.000000 | 11/13/ -14 | -18.360765 | -27.744403 |
| | 1716704.000000 | 01/30/ -13 | -18.327623 | -27.694077 |
| | 1716991.000000 | 11/12/ -13 | -18.297289 | -27.648026 |
| | 1718165.000000 | 01/30/ -9 | -18.320064 | -27.682601 |
| | 1718452.000000 | 11/12/ -9 | -18.304491 | -27.658959 |
| | 1719626.000000 | 01/30/ -5 | -18.310559 | -27.668170 |
| | 1719913.000000 | 11/12/ -5 | -18.313476 | -27.672599 |
| | 1721087.000000 | 01/30/ -1 | -18.300947 | -27.653578 |
| | 1721374.000000 | 11/12/ -1 | -18.322035 | -27.685592 |
| | 1721452.000000 | 01/29/ 1 | -18.363730 | -27.748906 |
| | 1722548.000000 | 01/30/ 4 | -18.293230 | -27.641865 |
| | 1722835.000000 | 11/12/ 4 | -18.328618 | -27.695588 |

| h=1,12° | 1711956.000000 | 01/30/ -26 | -18.290132 | -27.648581 |
|---|---|---|---|---|
| | 1712243.000000 | 11/13/ -26 | -18.337816 | -27.720980 |
| | 1712321.000000 | 01/30/ -25 | -18.352704 | -27.743590 |
| | 1712608.000000 | 11/12/ -25 | -18.274982 | -27.625584 |
| | 1713417.000000 | 01/30/ -22 | -18.280233 | -27.633554 |
| | 1713704.000000 | 11/13/ -22 | -18.346936 | -27.734830 |
| | 1713782.000000 | 01/30/ -21 | -18.342843 | -27.728613 |
| | 1714069.000000 | 11/12/ -21 | -18.284016 | -27.639297 |
| | 1714878.000000 | 01/30/ -18 | -18.271264 | -27.619942 |
| | 1715165.000000 | 11/13/ -18 | -18.354696 | -27.746615 |
| | 1715243.000000 | 01/30/ -17 | -18.334429 | -27.715836 |
| | 1715530.000000 | 11/12/ -17 | -18.291286 | -27.650333 |
| | 1716704.000000 | 01/30/ -13 | -18.327623 | -27.705501 |
| | 1716991.000000 | 11/12/ -13 | -18.297289 | -27.659446 |
| | 1718165.000000 | 01/30/ -9 | -18.320064 | -27.694024 |
| | 1718452.000000 | 11/12/ -9 | -18.304491 | -27.670379 |
| | 1719626.000000 | 01/30/ -5 | -18.310559 | -27.679591 |
| | 1719913.000000 | 11/12/ -5 | -18.313476 | -27.684021 |
| | 1721087.000000 | 01/30/ -1 | -18.300947 | -27.664998 |
| | 1721374.000000 | 11/12/ -1 | -18.322035 | -27.697015 |
| | 1722548.000000 | 01/30/ 4 | -18.293230 | -27.653284 |
| | 1722835.000000 | 11/12/ 4 | -18.328618 | -27.707012 |
| h=1,13 ° | 1711956.000000 | 01/30/ -26 | -18.290132 | -27.660001 |
| | 1712243.000000 | 11/13/ -26 | -18.337816 | -27.732408 |
| | 1712608.000000 | 11/12/ -25 | -18.274982 | -27.637001 |
| | 1713417.000000 | 01/30/ -22 | -18.280233 | -27.644973 |
| | 1713704.000000 | 11/13/ -22 | -18.346936 | -27.746260 |
| | 1713782.000000 | 01/30/ -21 | -18.342843 | -27.740042 |
| | 1714069.000000 | 11/12/ -21 | -18.284016 | -27.650716 |
| | 1714878.000000 | 01/30/ -18 | -18.271264 | -27.631359 |
| | 1715243.000000 | 01/30/ -17 | -18.334429 | -27.727263 |
| | 1715530.000000 | 11/12/ -17 | -18.291286 | -27.661754 |
| | 1716339.000000 | 01/30/ -14 | -18.264196 | -27.620629 |
| | 1716704.000000 | 01/30/ -13 | -18.327623 | -27.716928 |
| | 1716991.000000 | 11/12/ -13 | -18.297289 | -27.670867 |
| | 1718165.000000 | 01/30/ -9 | -18.320064 | -27.705449 |
| | 1718452.000000 | 11/12/ -9 | -18.304491 | -27.681802 |
| | 1719626.000000 | 01/30/ -5 | -18.310559 | -27.691015 |
| | 1719913.000000 | 11/12/ -5 | -18.313476 | -27.695445 |
| | 1721087.000000 | 01/30/ -1 | -18.300947 | -27.676420 |
| | 1721374.000000 | 11/12/ -1 | -18.322035 | -27.708441 |
| | 1721739.000000 | 11/12/ 1 | -18.258687 | -27.612268 |
| | 1722548.000000 | 01/30/ 4 | -18.293230 | -27.664704 |
| | 1722835.000000 | 11/12/ 4 | -18.328618 | -27.718439 |

| h=1,14 ° | 1711956.000000 | 01/30/ -26 | -18.290132 | -27.671424 |
| --- | --- | --- | --- | --- |
| | 1712243.000000 | 11/13/ -26 | -18.337816 | -27.743838 |
| | 1712608.000000 | 11/12/ -25 | -18.274982 | -27.648421 |
| | 1713417.000000 | 01/30/ -22 | -18.280233 | -27.656393 |
| | 1714069.000000 | 11/12/ -21 | -18.284016 | -27.662137 |
| | 1714878.000000 | 01/30/ -18 | -18.271264 | -27.642778 |
| | 1715243.000000 | 01/30/ -17 | -18.334429 | -27.738693 |
| | 1715530.000000 | 11/12/ -17 | -18.291286 | -27.673176 |
| | 1716339.000000 | 01/30/ -14 | -18.264196 | -27.632047 |
| | 1716704.000000 | 01/30/ -13 | -18.327623 | -27.728356 |
| | 1716991.000000 | 11/12/ -13 | -18.297289 | -27.682290 |
| | 1717800.000000 | 01/30/ -10 | -18.257019 | -27.621153 |
| | 1718165.000000 | 01/30/ -9 | -18.320064 | -27.716876 |
| | 1718452.000000 | 11/12/ -9 | -18.304491 | -27.693226 |
| | 1719626.000000 | 01/30/ -5 | -18.310559 | -27.702440 |
| | 1719913.000000 | 11/12/ -5 | -18.313476 | -27.706871 |
| | 1720278.000000 | 11/12/ -4 | -18.250547 | -27.611330 |
| | 1721087.000000 | 01/30/ -1 | -18.300947 | -27.687844 |
| | 1721374.000000 | 11/12/ -1 | -18.322035 | -27.719868 |
| | 1721739.000000 | 11/12/ 1 | -18.258687 | -27.623685 |
| | 1722548.000000 | 01/30/ 4 | -18.293230 | -27.676127 |
| | 1722835.000000 | 11/12/ 4 | -18.328618 | -27.729868 |
| h=1,15 ° | 1711956.000000 | 01/30/ -26 | -18.290132 | -27.682848 |
| | 1712608.000000 | 11/12/ -25 | -18.274982 | -27.659843 |
| | 1713417.000000 | 01/30/ -22 | -18.280233 | -27.667816 |
| | 1714069.000000 | 11/12/ -21 | -18.284016 | -27.673561 |
| | 1714878.000000 | 01/30/ -18 | -18.271264 | -27.654199 |
| | 1715530.000000 | 11/12/ -17 | -18.291286 | -27.684601 |
| | 1716339.000000 | 01/30/ -14 | -18.264196 | -27.643468 |
| | 1716704.000000 | 01/30/ -13 | -18.327623 | -27.739787 |
| | 1716991.000000 | 11/12/ -13 | -18.297289 | -27.693716 |
| | 1717800.000000 | 01/30/ -10 | -18.257019 | -27.632572 |
| | 1718165.000000 | 01/30/ -9 | -18.320064 | -27.728306 |
| | 1718452.000000 | 11/12/ -9 | -18.304491 | -27.704653 |
| | 1719261.000000 | 01/30/ -6 | -18.247885 | -27.618706 |
| | 1719626.000000 | 01/30/ -5 | -18.310559 | -27.713868 |
| | 1719913.000000 | 11/12/ -5 | -18.313476 | -27.718299 |
| | 1720278.000000 | 11/12/ -4 | -18.250547 | -27.622748 |
| | 1721087.000000 | 01/30/ -1 | -18.300947 | -27.699270 |
| | 1721374.000000 | 11/12/ -1 | -18.322035 | -27.731298 |
| | 1721739.000000 | 11/12/ 1 | -18.258687 | -27.635105 |
| | 1722548.000000 | 01/30/ 4 | -18.293230 | -27.687552 |
| | 1722835.000000 | 11/12/ 4 | -18.328618 | -27.741298 |
| h=1,16 ° | 1711956.000000 | 01/30/ -26 | -18.290132 | -27.694275 |
| | 1712608.000000 | 11/12/ -25 | -18.274982 | -27.671268 |
| | 1713417.000000 | 01/30/ -22 | -18.280233 | -27.679241 |
| | 1714069.000000 | 11/12/ -21 | -18.284016 | -27.684986 |
| | 1714878.000000 | 01/30/ -18 | -18.271264 | -27.665623 |
| | 1715530.000000 | 11/12/ -17 | -18.291286 | -27.696028 |
| | 1716339.000000 | 01/30/ -14 | -18.264196 | -27.654890 |
| | 1716991.000000 | 11/12/ -13 | -18.297289 | -27.705144 |
| | 1717800.000000 | 01/30/ -10 | -18.257019 | -27.643993 |
| | 1718165.000000 | 01/30/ -9 | -18.320064 | -27.739737 |
| | 1718452.000000 | 11/12/ -9 | -18.304491 | -27.716082 |
| | 1718817.000000 | 11/12/ -8 | -18.241363 | -27.620225 |
| | 1719261.000000 | 01/30/ -6 | -18.247885 | -27.630126 |
| | 1719626.000000 | 01/30/ -5 | -18.310559 | -27.725298 |
| | 1719913.000000 | 11/12/ -5 | -18.313476 | -27.729730 |
| | 1720278.000000 | 11/12/ -4 | -18.250547 | -27.634168 |
| | 1720722.000000 | 01/30/ -2 | -18.237995 | -27.615113 |
| | 1721087.000000 | 01/30/ -1 | -18.300947 | -27.710699 |
| | 1721374.000000 | 11/12/ -1 | -18.322035 | -27.742730 |
| | 1721739.000000 | 11/12/ 1 | -18.258687 | -27.646526 |
| | 1722548.000000 | 01/30/ 4 | -18.293230 | -27.698979 |
| h=1,17 ° | 1711591.000000 | 01/30/ -27 | -18.227149 | -27.610067 |
| | 1711956.000000 | 01/30/ -26 | -18.290132 | -27.705704 |
| | 1712608.000000 | 11/12/ -25 | -18.274982 | -27.682694 |
| | 1713417.000000 | 01/30/ -22 | -18.280233 | -27.690669 |
| | 1714069.000000 | 11/12/ -21 | -18.284016 | -27.696414 |
| | 1714878.000000 | 01/30/ -18 | -18.271264 | -27.677049 |
| | 1715530.000000 | 11/12/ -17 | -18.291286 | -27.707457 |
| | 1715895.000000 | 11/12/ -16 | -18.227477 | -27.610566 |
| | 1716339.000000 | 01/30/ -14 | -18.264196 | -27.666315 |
| | 1716991.000000 | 11/12/ -13 | -18.297289 | -27.716574 |
| | 1717356.000000 | 11/12/ -12 | -18.233591 | -27.619847 |
| | 1717800.000000 | 01/30/ -10 | -18.257019 | -27.655416 |
| | 1718452.000000 | 11/12/ -9 | -18.304491 | -27.727514 |
| | 1718817.000000 | 11/12/ -8 | -18.241363 | -27.631646 |
| | 1719261.000000 | 01/30/ -6 | -18.247885 | -27.641548 |
| | 1719626.000000 | 01/30/ -5 | -18.310559 | -27.736730 |
| | 1719913.000000 | 11/12/ -5 | -18.313476 | -27.741162 |
| | 1720278.000000 | 11/12/ -4 | -18.250547 | -27.645591 |
| | 1720722.000000 | 01/30/ -2 | -18.237995 | -27.626533 |
| | 1721087.000000 | 01/30/ -1 | -18.300947 | -27.722129 |
| | 1721739.000000 | 11/12/ 1 | -18.258687 | -27.657950 |
| | 1722183.000000 | 01/30/ 3 | -18.229726 | -27.613980 |
| | 1722548.000000 | 01/30/ 4 | -18.293230 | -27.710409 |
| h=1,18 ° | 1711591.000000 | 01/30/ -27 | -18.227149 | -27.621488 |
| | 1711956.000000 | 01/30/ -26 | -18.290132 | -27.717135 |
| | 1712608.000000 | 11/12/ -25 | -18.274982 | -27.694122 |
| | 1713417.000000 | 01/30/ -22 | -18.280233 | -27.702098 |
| | 1714069.000000 | 11/12/ -21 | -18.284016 | -27.707844 |
| | 1714434.000000 | 11/12/ -20 | -18.220665 | -27.611645 |
| | 1714878.000000 | 01/30/ -18 | -18.271264 | -27.688477 |
| | 1715530.000000 | 11/12/ -17 | -18.291286 | -27.718888 |
| | 1715895.000000 | 11/12/ -16 | -18.227477 | -27.621986 |
| | 1716339.000000 | 01/30/ -14 | -18.264196 | -27.677741 |
| | 1716991.000000 | 11/12/ -13 | -18.297289 | -27.728006 |
| | 1717356.000000 | 11/12/ -12 | -18.233591 | -27.631269 |
| | 1717800.000000 | 01/30/ -10 | -18.257019 | -27.666842 |
| | 1718452.000000 | 11/12/ -9 | -18.304491 | -27.738947 |
| | 1718817.000000 | 11/12/ -8 | -18.241363 | -27.643069 |
| | 1719261.000000 | 01/30/ -6 | -18.247885 | -27.652972 |
| | 1719626.000000 | 01/30/ -5 | -18.310559 | -27.748165 |
| | 1720278.000000 | 11/12/ -4 | -18.250547 | -27.657015 |
| | 1720722.000000 | 01/30/ -2 | -18.237995 | -27.637956 |
| | 1721087.000000 | 01/30/ -1 | -18.300947 | -27.733562 |
| | 1721739.000000 | 11/12/ 1 | -18.258687 | -27.669376 |
| | 1722183.000000 | 01/30/ 3 | -18.229726 | -27.625401 |
| | 1722548.000000 | 01/30/ 4 | -18.293230 | -27.721840 |

**Tavola β**

**Le date di Augusto**

| | |
|---|---|
| 63 a.C. 23 settembre | Nascita di Gaio Ottavio |
| 31 a.C. 2 settembre | Battaglia di Azio |
| 30 a.C. 1 agosto | Presa di Alessandria e suicidio di Cleopatra |
| 29 a.C. 13–15 agosto | Triplice trionfo di Ottaviano |
| 29 a.C. 18 agosto | Dedica del tempio del *Divus Iulius* nel Foro Romano |
| 28 a.C. 9 ottobre | Dedica del Tempio di Apollo sul Palatino |
| 27 a.C. 13 gennaio | Augusto restituisce al senato i poteri magistraturali |
| 27 a.C. 16 gennaio | Ottaviano ottiene dal Senato il titolo di *Augustus* |
| 19 a.C. 21 settembre | Morte di Virgilio e pubblicazione dell'Eneide |
| 15 a.C. 24 maggio | Nascita di Germanico |
| 13 a.C. 4 luglio | Al ritorno di Augusto dalla Spagna, il Senato decreta la costruzione dell'*ara Pacis Augustae* |
| 12 a.C. 6 marzo | Augusto assume la carica di pontefice massimo |
| 12 a.C. 28 aprile | Dedica di un *signum Vestae* sul Palatino. |
| 9 a.C. 30 gennaio | Dedica dell'*Ara Pacis Augustae*. |
| 9 a.C. 14 settembre | Muore Druso in Germania |
| 9 a.C. 9–11 settembre | Sconfitta di Varo |
| 14 d.C. 19 agosto | Augusto muore a Nola |
| 14 d.C. 17 settembre | Il Senato proclama Augusto *divus* |